\documentclass[notitlepage,12pt,nofootinbib,tightenlines,prb,superscript address]{revtex4-2}
\usepackage{epsfig}
\usepackage{array}
\usepackage{color}
\usepackage{bm}
\usepackage{todonotes}
\usepackage[nottoc,notlof,notlot]{tocbibind}
\usepackage[titles]{tocloft}

\usepackage{amsmath}
\usepackage{amssymb}
\usepackage{amsfonts}
\usepackage{graphicx}
\usepackage{cancel}
\usepackage{fullpage}
\usepackage[colorlinks=true,
            linkcolor=blue,
            urlcolor=blue,
            citecolor=blue]{hyperref}
\usepackage{setspace}
\usepackage{etoolbox}
\usepackage{diagbox}
\usepackage{booktabs}

\newcommand{\Lagr}{\mathcal{L}}

\newcommand{\llangle}{\langle\!\langle}
\newcommand{\rrangle}{\rangle\!\rangle}

\newcommand{\non}{\nonumber\\}

\newcommand{\beq}{\begin{equation}}
	\newcommand{\eeq}{\end{equation}}
\newcommand{\beqn}{\begin{eqnarray}}
	\newcommand{\eeqn}{\end{eqnarray}}

\newcommand{\gsim}{\lower.7ex\hbox{$
		\;\stackrel{\textstyle>}{\sim}\;$}}
\newcommand{\lsim}{\lower.7ex\hbox{$
		\;\stackrel{\textstyle<}{\sim}\;$}}
\begin{document}
\title{Skyrmions and Hopfions in 3D Frustrated Magnets}
\author{Carlos Naya}
\affiliation{Department of Physics, Stockholm University, AlbaNova University Center, 106 91 Stockholm, Sweden}
\author{Daniel Schubring}
\affiliation{School of Physics and Astronomy, University of Minnesota, Minneapolis, MN 55455, USA}
\author{Mikhail Shifman}
\affiliation{School of Physics and Astronomy, University of Minnesota, Minneapolis, MN 55455, USA}
\affiliation{William I. Fine Theoretical Physics Institute, University of Minnesota, Minneapolis, MN 554455, USA}
\author{Zhentao Wang}
\affiliation{School of Physics and Astronomy, University of Minnesota, Minneapolis, MN 55455, USA}
\begin{abstract}
	A model of an inversion-symmetric frustrated spin system is introduced which hosts three-dimensional extensions of magnetic Skyrmions. In the continuum approximation this model reduces to a non-linear sigma model on a squashed sphere which has a natural interpolating parameter. At one limit of the parameter the model reduces to a frustrated magnetic system earlier considered by Sutcliffe as a host to Hopfions, and in the other limit it becomes very similar to the 3D Skyrme model. To better understand the relation between Hopfions and 3D Skyrmions a model interpolating between the Faddeev-Niemi model and the Skyrme model is reconsidered and it is shown that energies of the solitons obey a linear BPS bound. The 3D Skyrmions in the frustrated magnetic model are found and compared to the rational map ansatz.
\end{abstract}
	\maketitle
	\tableofcontents
%\begin{spacing}{1.25}
	\section{Introduction}
	The occurrence of solitons was first described in the literature as early as 1834 by John Scott Russell. Theoretical studies of topologically stable solitons began  in earnest in the mid-1960s,
	see e.g.\!\! the collection~\cite{soliani1984solitons}. The discovery of instantons~\cite{BPST} and 't Hooft-Polyakov monopoles~\cite{hooft1974magnetic,polyakov1974particle} after the advent of Yang-Mills theories provided an impetus to further rapid advances in this area.

In this paper we will investigate the close interrelationship of several notions of topological charge in systems with three spatial dimensions. In particular we will focus on a lattice spin system which supports three-dimensional Skyrmions in the original sense of Skyrme~\cite{skyrme,WITTEN1983433,adkins1983static} and connect these to Hopfions~\cite{faddeev1997stable} (reviewed in~\cite{radu2008stationary}) which are widely discussed as a generalization of the two-dimensional magnetic Skyrmions (also known as baby Skyrmions) which are more familiar in a condensed matter context (see e.g.~\cite{Nagaosa2013,han2017skyrmions,Back2020}).

While much work on magnetic Skyrmions involves chiral ferromagnets with a Dzyaloshinskii-Moriya interaction~\cite{DzMo}, more recently there has been a lot of interest in topological spin textures in inversion-symmetric magnets~\cite{okubo2012multiple,leonov2015multiply,lin2016ginzburg,Wang2020} where the magnetic Skyrmions are stabilized by competing magnetic interactions. The three-dimensional model we will consider here is a frustrated spin system of this type, and in one limit is very close to the effective theory studied by Sutcliffe~\cite{sutcliffe2017skyrmion} as a medium for Hopfions.

A Hopfion is a topological defect which may be considered to be a magnetic Skyrmion extended in the third dimension to form a loop of string. Such magnetic Skyrmion strings have been observed experimentally~\cite{milde2013unwinding,seki2021direct}, and moreover Hopfions themselves have recently been constructed in magnetic systems~\cite{kent2021creation}. {There has also been much discussion of Hopfions in two-component superconductors \cite{babaev2002hidden,gorsky2013revisiting}, which although in the simplest case are not energetically stable \cite{jaykka2011supercurrent} they may perhaps be stabilized by a current-current interaction \cite{rybakov2019stable}.} The topological stability of such Hopfions is provided for by the Hopf invariant~\cite{whitehead1947expression} which is closely analogous to the notion of helicity in hydrodynamic systems~\cite{kamchatnov1982topological,moffatt2014helicity}, and in this sense knotted `Hopfions' have even been constructed experimentally in fluids~\cite{kleckner2013creation}. 

Such knotted topological solitons have captured the interest of many working with the model of Faddeev and Niemi~\cite{faddeev1997stable,faddeev1976some,faddeev1997toroidal,battye1998knots,sutcliffe2007knots}\footnote{{This model is often referred to as the Skyrme-Faddeev model, but in order to more clearly distinguish it from the Skyrme model proper we will refer to it as the \emph{Faddeev-Niemi} model.}}, and in such papers attention is often brought to the old idea of Lord Kelvin suggesting knotted vortices might be related to elementary particles~\cite{LordKelvin}. Of course there is a different but related model involving topological defects which really is believed to have some connection to baryons, and that is the \emph{Skyrme model}~\cite{skyrme,WITTEN1983433,adkins1983static} (see also the review~\cite{makhan1992skyrme}).

Skyrmions proper in the sense of the three-dimensional {(3D)} Skyrme model are not as familiar in a condensed matter context, but
{they have} been predicted to emerge near the Lifshitz point (commensurate to incommensurate transition) of 3D magnetic systems with non-collinear ground states~\cite{BatistaEtAl2018}.
For certain choices of parameters,
the {second-derivative terms of the} effective field theory of
{such frustrated lattice models are} equivalent to the $SU(2)$ principal chiral model which is a major component of the Skyrme model. The simplified lattice model considered here
{can be regarded as the low-energy effective model for a broad class of 3D frustrated magnets with non-collinear ground states,}
and it too reduces to the principal chiral model in one limit of parameters. While the higher order terms in the effective field theory expansion of the frustrated lattice models differ from the Skyrme term, they too
{were predicted to}
stabilize 3D Skyrmions {near the Lifshitz point}~\cite{BatistaEtAl2018}, as will be shown here explicitly.

 The lattice model considered here involves multiple spins at each lattice site of a cubic lattice, much like how a pyrochlore lattice involves four spins at each tetrahedral cell of an outer face-centered-cubic lattice.  Large scale rigid rotations of these spins lead to $SO(3)$ Goldstone modes much as in the principal chiral model. Typically the low-energy theory of realistic frustrated magnetic models involves both these rigid rotations as well as other modes that modify the relative angles between neighboring spins. However these extra modes can be gapped out by certain interactions, for instance the biquadratic term on the pyrochlore lattice model~\cite{BatistaEtAl2018}. So to provide a low-energy effective model for a large class of realistic 3D non-collinear magnets we may simply take the relative angles between spins on the same site to be fixed. A key idea used here is that when the relative angle between spins takes a special value then the effective description is much like the Skyrme model, i.e. a non-linear sigma model on a three-sphere $S^3$. As the vacuum configuration of spins becomes more and more colinear the target space deforms to a \emph{squashed sphere} \cite{azaria1995massive,squashedsigma}, and in the limit of perfect colinearity the model becomes equivalent to a non-linear sigma model with an $S^2$ target space much as is in the Faddeev-Niemi model or the frustrated magnetic models previously considered as a host to Hopfions \cite{sutcliffe2017skyrmion}.

A very similar continuum model deforming the Skyrme model to the Faddeev-Niemi model has been considered previously by Nasir and Niemi~\cite{nasir2002effective} and Ward and Silva Lobo~\cite{ward2004skyrmions,lobo2011generalized,silva2011lattices}. We will also reconsider this model here as a close analogy to the lattice model, and show that it obeys a linear Bogomol’nyi-Prasad-Sommerfeld (BPS)~\cite{BPS} bound on the energy. Note however that the connection between 3D Skyrmions and Hopfions in these models is entirely different from another notion of Hopfions in the Skyrme model pointed out by Meissner~\cite{meissner1985toroidal} and Cho~\cite{cho2001monopoles,cho2008new}. In our case we wish to stress that the natural projection map from $S^3$ to $S^2$ implies that 3D Skyrmions themselves may be considered as Hopfions and vice versa. This is an idea that occasionally appears in the literature and in particular underlies Ward's treatment of the continuum model~\cite{ward2004skyrmions}. The squashing of the sphere changes the energy functional and thus the quantitative features of the minimum energy soliton, but there is no dramatic qualitative difference between the topological defects of the Skyrme model and the Faddeev-Niemi model.

The analogue of this statement will also be shown explicitly in numerical simulations of the frustrated magnetic model which we introduce here. First it will be shown that the model has 3D Skyrmion solutions much like the Skyrme model. The unit Skyrmion will be seen to be quantitatively very close to the spherically symmetric hedgehog solution, and small clusters of Skyrmions may be approximated by the same rational map ansatz used for both the Skyrme model and BPS monopoles in $SU(2)$ Yang-Mills~\cite{houghton1998rational,battye2002skyrmions}. However as the charge increases the Skyrmions in the lattice model will exhibit new shapes departing from the rational map ansatz. And as the squashing parameter of the model increases the Skyrmion clusters, which are similar in some respects to models of nuclei, will be shown to deform to twisted{, linked} or knotted strings, much like a modern incarnation of Lord Kelvin's idea.
\subsection{Outline}

This paper is divided into two main sections. Sec.~\ref{sec 2} deals with the effective theory of the squashed sphere sigma model and it introduces the notation and necessary topological concepts in that context in Sec.~\ref{sec Squashed sphere} and \ref{sec Top charge} respectively. The continuum model of Nasir, Niemi~\cite{nasir2002effective} and Ward~\cite{ward2004skyrmions} interpolating between the Skyrme model and the Faddeev-Niemi model is reconsidered with some new numerical simulations and a new theoretical result on energy bounds in Sec.~\ref{sec Ward model}. Finally the notion of the equivalence between 3D Skyrmions and Hopfions is discussed a bit further in Sec.~\ref{sec Position curves and strings}, where an ansatz for Skyrmion strings which have baryon charge per length is also introduced and compared to previous results in the Faddeev-Niemi model.

Then in Sec.~\ref{sec 3} the main frustrated magnetic system is considered. Most of the discussion in Sec.~\ref{sec 2} will be applicable to the continuum description of this model as well. The lattice model is introduced in Sec.~\ref{sec Lattice model intro} and its continuum description is found in Sec.~\ref{sec Lattice model effective theory}. The details of the numerical simulation are introduced in Sec.~\ref{sec Lattice model numerical results} and results on the unit charge Skyrmion are compared to the hedgehog ansatz in the continuum description. In Sec.~\ref{sec Rational map} higher charge Skyrmion configurations are considered and compared to the rational map ansatz in the continuum description.
%For charges less than or equal to four the rational map ansatz is a reasonably good description of minimum energy soliton, but for higher charges the solution is shown to be quite different.
{In Sec.~\ref{sec charge-10} we show numerical results for the charge-10 Skyrmion, with emphasis on the interpolation of the topological charge isosurfaces and the position curves. } 
Finally, in the concluding Sec.~\ref{sec Conclusion} the possible connection to experiment and further investigation of the theoretical model in terms of Skyrmion lattices is discussed.

%Some preliminary results on the variation of higher charge configurations with the squashing parameter are presented for both the model of Ward and the lattice model in Sec.~\ref{sec Position curves and strings} and Sec.~\ref{sec Rational map} respectively and more will be included in a future update of this preprint.
%\zw{[xx This sentence to be removed xx]}

\section{Squashing the Skyrme model}\label{sec 2}%Instantons, Hopfions, and Skyrmions

Since our aim is to introduce a model which is closely related to both the Skyrme model and the Faddeev-Niemi model, let us begin by reviewing these continuum models and illustrating the connection between them. The notation and discussion on topological charge in this context will be directly applicable to the lattice model which is our main focus in the next section.

\subsection{The squashed sphere non-linear sigma model} \label{sec Squashed sphere}

	The terms of the Skyrme model which are quadratic in derivatives are identical to the $SU(2)$ principal chiral model (PCM), which is expressed in terms of a matrix field $U\in SU(2)$ and a parameter $f_\pi$ with dimensions of  energy,
	\begin{align}
		\Lagr_{PCM}=\frac{f_\pi^2}{4}\text{Tr}\left(\partial_\mu U^{-1}\partial_\mu U\right).\label{lagr PCM U}
	\end{align}
This has a global symmetry under right multiplication $U\rightarrow UV_R$, and the three independent Noether currents $J^i_\mu$ corresponding to this symmetry are
\begin{align}
	J^i_\mu = \frac{1}{2}\text{Tr}\left(J_\mu \sigma^i\right),\qquad J_\mu \equiv -i U^{-1}\partial_\mu U,\label{def J currents}
\end{align}
where $\sigma^i$ are the standard Pauli matrices with normalization $\text{Tr}\left(\sigma^i\sigma^j\right)=2\delta^{ij}$. Momentarily we will consider models where the global symmetry associated with $J$ is explicitly broken to a $U(1)$ subgroup (although the full global symmetry under \emph{left} multiplication will be maintained) but these quantities $J$ will still be very useful, and the Lagrangian may be expressed in terms of them,
\begin{align}
	\Lagr_{PCM}=\frac{f_\pi^2}{2}\sum_{i=1,2,3}\left(J^i_\mu\right)^2.\label{lagr PCM J}
\end{align}

On the other hand, we also wish to consider the Faddeev-Niemi model which involves a three-component real unit vector $\left(S^i\right)^2=1$, and the quadratic terms in the action are just that of the $O(3)$ non-linear sigma model. As usual, the action may instead be expressed in the form of a complex two-component unit vector $z^\alpha$, which is connected to real unit vector $S^i$ through the Pauli matrices $\sigma^i$,
\begin{align}
	S^i\equiv -\bar{z}^\alpha \sigma^i_{\alpha\beta}z^\beta, \qquad \bar{z}^\alpha z^\alpha=1.\label{def S and z}
	\end{align}
This change of fields leads to the $CP^1$ form of the non-linear sigma model,
\begin{align}
	\Lagr_{CP^1}=\frac{f_\pi^2}{2}\left(\partial \bar{z}\cdot\partial z +\left(\bar{z}\cdot\partial z\right)^2\right)=\frac{f_\pi^2}{8}\left(\partial {S}\right)^2, \label{lagr CP1 S and z}
\end{align}
where the indices will be suppressed where obvious, and a dot may be used to clarify contraction of internal indices.

The $CP^1$ model above may be related to the PCM by expressing the action in terms of the special unitary matrix $U$ which is uniquely determined by $z$,
\begin{align}
	U=	\left(\begin{array}{cc}
			\bar{z}^1& z^0\\
		-\bar{z}^0 & z^1
		\end{array}\right), \label{def U and z}
\end{align}
and then further in terms of the $J$ currents defined above in \eqref{def J currents},
\begin{align}
	\Lagr_{CP^1}=\frac{f_\pi^2}{2}\sum_{a=1,2}\left(J^a_\mu\right)^2. \label{lagr CP1 J}
\end{align}
The only difference from the PCM case \eqref{lagr PCM J} is that the sum only runs over two components. To better distinguish the two cases even in the absence of an explicit summation symbol, a Latin index from the beginning of the alphabet will run over $1,2$, and a Latin index from the middle of the alphabet will run over $1,2,3$.

In this form there is an obvious interpolation between the PCM and $CP^1$ model which may be constructed in terms of a parameter $\beta$ ranging from $0$ to $1$, respectively\footnote{$\beta$ may also be continued to negative values, which is relevant in e.g. \cite{dombre1989nonlinear}.},
\begin{align}
	\Lagr_\beta = \frac{f_\pi^2}{2}\left[\left(J^i_\mu\right)^2-\beta\left(J^3_\mu\right)^2\right]. \label{lagr Squashed J}
\end{align}
This Lagrangian actually has a clear geometric interpretation as a non-linear sigma model with a target space which is a \emph{squashed sphere} homeomorphic to $S^3$ but with a less symmetric metric. Such a model (with a specific negative value of $\beta$) has been shown by Dombre and Read to arise in 2D as an effective theory of a Heisenberg antiferromagnet on a triangular lattice~\cite{dombre1989nonlinear}, and the renormalization of the continuum model has previously been considered in detail~\cite{azaria1995massive,squashedsigma}. The 3D frustrated magnetic model introduced below in Sec.~\ref{sec 3} was specifically chosen to produce this squashed sphere sigma model action in the continuum limit much as was done by Dombre and Read in the 2D case. As we will show there, terms involving higher order derivatives will naturally arise in the continuum approximation to the lattice model, and these terms will allow for the presence of stable topological defects.

	\subsection{Gauge symmetry and topological charges}\label{sec Top charge}

	In the $\beta=1$ limit, the model reduces to the $CP^1$ model \eqref{lagr CP1 S and z} which has a $U(1)$ gauge symmetry under transformations $z(x)\rightarrow e^{-i\phi(x)}z(x)$, where $\phi(x)$ is an arbitrary function of the spatial coordinate $x$. In terms of $U$ in \eqref{def U and z}, this gauge symmetry corresponds to right multiplication by the unitary matrix $V_R=\text{diag}\left(e^{i\phi(x)}, e^{-i\phi(x)}\right)$, from which it is easily shown that the $J$ currents transform as,
	\begin{align}J^3\rightarrow J^3+\partial \phi,\qquad \left(\begin{array}{c}
	J^1 \\
	J^2
\end{array}\right)\rightarrow\left(\begin{array}{cc}
	\cos 2\phi & -\sin 2\phi\\
	\sin 2\phi & \cos 2\phi
\end{array}\right)\left(\begin{array}{c}
	J^1 \\
	J^2
\end{array}\right).\label{eq J gauge transf}
	\end{align}

Clearly the structure in the $CP^1$ action, $\left(J^1\right)^2+\left(J^2\right)^2$, is gauge invariant since there is no explicit $J^3$ dependence and the quadratic form is invariant under rotations of $J^1, J^2$. There are two other obvious gauge invariant structures that may be constructed. The two-form $J^1_\mu J^2_\nu-J^2_\mu J^1_\nu$ is also invariant under rotations of $J^1, J^2$. And given that $J^3$ transforms like a vector potential
\begin{align}
	A_\mu\equiv J^3_\mu,\label{def A}
	\end{align}
the gauge invariant field strength tensor $F_{\mu\nu}=\partial_\mu A_\nu-\partial_\nu A_\mu$ may also be constructed. In fact by expressing the $J$ currents in terms of the $z$ field it can be quickly shown that these two quantities are not independent,
\begin{align}
		F_{\mu\nu}\equiv \partial_\mu J^3_\nu -\partial_\nu J^3_\mu= 2\left(J^1_\mu J^2_\nu -J^2_\mu J^1_\nu\right).\label{def F tensor}
\end{align}

Just as the gauge invariant $CP^1$ action \eqref{lagr CP1 S and z} may be written entirely in terms of the real unit vector field $S^i$, so may the gauge invariant $F_{\mu\nu}$ tensor.\footnote{This equality can be shown by exploiting the global symmetry to choose $S^i=(0,0,1)$ and expressing $F$ in terms of the complex field $z^\alpha=(z^0,0)$.}
\begin{align}
	F_{\mu\nu}=\frac{1}{2}\epsilon_{ijk}S^i\partial_\mu S^j \partial_\nu S^k.\label{eq F tensor S}
\end{align}

This $F$ tensor is directly related to the notion of topological charge for two-dimensional magnetic (baby) Skyrmion field configurations.\footnote{In a high-energy context this same topological charge might be referred to as the instanton charge of the 2D $CP^1$ model.} As a two-form, $F$ may be integrated over an arbitrary two-dimensional surface $\Sigma$, and the result will be the magnetic Skyrmion charge counting the number of times the map $S^i:\Sigma\rightarrow S^2$ wraps around the $S^2$ target space of the $S^i$ field (up to a $2\pi$ difference in normalization).

Furthermore, $F$ may be used to define a $U(1)$ Chern-Simons three-form $A\wedge F$ which represents a distinct notion of topological charge density which is integrated over 3D volume rather than a 2D surface. This is just the Hopf charge $Q$~\cite{whitehead1947expression},
\begin{align}
	Q= -\frac{1}{8\pi^2}\int d^3x \,\epsilon^{\lambda\mu\nu}A_\lambda F_{\mu\nu}. \label{def Q Hopf}
\end{align}
Roughly speaking, a field configuration with non-zero Hopf charge may be described as a 2D magnetic Skyrmion (or $CP^1$ instanton) extended as a string in the third spatial direction, and then tied back in a loop. For this loop to be topologically distinct from the $Q=0$ vacuum it must be twisted or knotted in a non-trivial way~\cite{moffatt1995helicity}.

So far we have been considering two forms of topological charge which, due to the gauge invariance of the quantities involved, are able to be expressed in terms of the unit vector $S^i$ field which maps physical space to the $S^2$ target space. However, we began with the squashed sphere sigma model \eqref{lagr Squashed J} and the currents $J$ which are expressible in terms of the $U\in SU(2)$ field. Since $SU(2)$ is homeomorphic to the three sphere $S^3$, there is another seemingly distinct form of topological charge which describes the windings of the $S^3$ base space\footnote{Our base space, i.e. ordinary physical space, is $R^3$ but due to the boundary condition at infinity it may be considered topologically equivalent to $S^3$. This boundary condition must also be applied to the gauge field $A$. } around the $S^3$ target space. This is referred to as the Skyrme charge or baryon charge,
\begin{align}
	Q= -\frac{1}{2\pi^2}\int d^3x \,\epsilon^{\lambda\mu\nu}J^1_\lambda J^2_\mu J^3_\nu. \label{def Q Skyrme}
\end{align}
But as can be seen from a direct substitution of the definitions of $A$ and $F$ in \eqref{def A},\eqref{def F tensor}, this is actually identical to the Hopf charge! This is the main point that we wish to stress, \emph{a Hopfion may be considered to be a three-dimensional Skyrmion, and vice-versa}. The $z$ field description of a Skyrmion may be directly mapped to the $S$ field using \eqref{def S and z}, and the result will have Hopf charge equal to its original baryon charge. On the other hand, a Hopfion involves a map to a $S^2$ target space which may be lifted to a $S^3$ target space by identifying the $A$ field with the $J^3$ field and integrating. Although this construction is not unique since $A$ is only defined up to a gauge transformation, for any choice of $A$ the lifted map will have baryon charge equal to its original Hopf charge.

This notion of the equivalence between the baryon charge and Hopf charge is not a new idea, it is clearly discussed in~\cite{radu2008stationary} and~\cite{han2017skyrmions} for instance. The idea also underlies the model of Ward which will be discussed further in the next subsection. Note however that this is distinct from a completely different notion of Hopfions in the Skyrme model~\cite{meissner1985toroidal,cho2001monopoles,cho2008new}, where a field configuration $U$ is restricted to only take values in a subspace $S^2\subset SU(2)$. In that case since the $U$ field does not cover $SU(2)$ the baryon charge vanishes, but a different notion of Hopf charge\footnote{The $F$ tensor for this second notion of Hopf charge is defined in terms of the unit vector $n$ which is considered in Sec.~\ref{sec Rational map} rather than $S$.} may still be defined in terms of the $S^2$ subset.

Finally, let us briefly comment on a third way in which the charge $Q$ may be understood which is more familiar from Yang-Mills theory. The current $J_\mu$ in \eqref{def J currents} may also be understood as a non-Abelian gauge field associated to gauge symmetry under right multiplication by $SU(2)$ matrices. It is pure gauge and the non-Abelian field strength tensor vanishes. If we consider the Chern-Simons three-form associated to this \emph{non-Abelian} gauge symmetry (as opposed to the Abelian gauge symmetry involved in the definition of the Hopf charge), we have the charge
$$\mathcal{K}=-\frac{1}{16\pi^2}\int d^3 x \, \epsilon^{\lambda\mu\nu}\left(J^k_\lambda \partial_\mu J^k_\nu + \frac{1}{3}\epsilon_{ijk}J_\lambda^iJ_\mu^jJ_\nu^k\right).$$
Rewriting $\partial_{[\mu}J^k_{\nu]}$ as a product of two gauge fields in a manner similar to \eqref{def F tensor}\footnote{This relation may also be understood as arising from the vanishing of non-Abelian field strength tensor.} we see that this just reduces to the expression for the baryon charge \eqref{def Q Skyrme}, so $\mathcal{K}=Q$. This non-Abelian Chern-Simons charge $\mathcal{K}$ is interesting in 4D Yang-Mills theory because 4D instantons can be understood as interpolating between 3D vacua with different values of $\mathcal{K}$~\cite{jackiw1976vacuum,CALLAN1976334,shifman1994instantons}.

This perspective on the charge is well-illustrated by the construction of Atiyah and Manton, where Skyrmion configurations with non-zero $Q$ are generated from an initial trivial configuration by integrating over $SU(2)$ instantons~\cite{Atiyah1989,Atiyah1993}. This gives a reasonably good approximation to the minimal energy configuration, and further work by Sutcliffe explained the success of the Atiyah-Manton approximation. In~\cite{Sutcliffe2010}, a novel BPS model is derived from the pure Yang-Mills theory in one higher dimension, obtaining a Skyrme field coupled to an infinite tower of vector mesons. Interestingly, when all vector mesons are considered, the BPS property is fulfilled with a Skyrme field given exactly by the holonomy of the instanton. Nevertheless, the restriction to the lowest vector meson already improves the Skyrmion description of nuclei, with low binding energies and nuclear cluster structures arising~\cite{Naya2018}.

	\subsection{The squashed Skyrme model and energy bounds}\label{sec Ward model}
		So far we have discussed the terms in the squashed sphere sigma model which are quadratic in derivatives, but due to Derrick's theorem~\cite{derrickstheorem} higher order terms are necessary to stabilize the topological defects with non-zero $Q$ which were discussed above. The Lagrangian of the full Skyrme model~\cite{skyrme} is
		\begin{align}
\Lagr_{Skyrme}=\frac{f_\pi^2}{2}\left[\left(J^k_\lambda\right)^2 + \frac{1}{2M^2}\epsilon_{ijk}\epsilon_{klm}J_\mu^iJ_\nu^jJ_\mu^lJ_\nu^m\right],
		\end{align}
	where $M$ is some new dimensionful parameter often written as $ef_\pi$. Using \eqref{def Q Skyrme}, the energy may be written as,
	\begin{gather*}
		E_{Skyrme}=\int d^3x\,\Lagr_{Skyrme}=E_{BPS}|Q|+	\frac{f_\pi^2}{2}\int d^3x\,\left(J^k_\lambda \pm \frac{1}{2M}\epsilon_{ijk}\epsilon^{\lambda\mu\nu}J_\mu^iJ_\nu^j\right)^2,\non\qquad E_{BPS}\equiv 6\pi^2 \frac{f_\pi^2}{M}.
	\end{gather*}
This form of the Skyrme energy functional clearly shows the BPS bound $E\geq  |Q|E_{BPS}$. This expression for the energy functional may easily be generalized to the squashed sphere case,
\begin{align*}
E_{BPS}|Q|+	\frac{f_\pi^2}{2}\int d^3x\left[\left(J^a_\lambda \pm \frac{1}{2M}\epsilon_{ija}\epsilon^{\lambda\mu\nu}J_\mu^iJ_\nu^j\right)^2+(1-\beta)\left(J^3_\lambda \pm \frac{1}{2M(1-\beta)}\epsilon_{ij3}\epsilon^{\lambda\mu\nu}J_\mu^iJ_\nu^j\right)^2\right],
\end{align*}
where $a$ is only summed over $1,2$. Note that the $\epsilon_{ij3}J^iJ^j$ expression in the $\beta$ dependent term is proportional to the $F$ tensor \eqref{def F tensor} defined above. Expanding the squares leads to the Lagrangian,
\begin{align}
	\Lagr=\frac{f_\pi^2}{2}\left[\left(J^k_\lambda\right)^2 -\beta\left(J^3_\lambda\right)^2+ \frac{1}{2M^2}\epsilon_{ijk}\epsilon_{klm}J_\mu^iJ_\nu^jJ_\mu^lJ_\nu^m+\frac{\beta}{8M^2(1-\beta)}\left(F_{\mu\nu}\right)^2\right],\label{lagr Ward}
\end{align}
which also satisfies the BPS bound
\begin{align}
	E\geq 12\pi^2\frac{f^2_\pi}{2M}|Q|\equiv E_{BPS}|Q|. \label{eq BPS bound}
\end{align}
The new term quartic in derivatives is exactly that of the Faddeev-Niemi model, so squashing the target space of the Skyrme model while maintaining the BPS bound naturally leads to an interpolation between the Skyrme and Faddeev-Niemi models. This generalized Skyrme system was considered earlier by Nasir and Niemi~\cite{nasir2002effective} and Ward and Silva Lobo~\cite{ward2004skyrmions,lobo2011generalized,silva2011lattices}.

It may seem that there is a difficulty in extending to the limit $\beta=1$ due to the prefactor $(1-\beta)^{-1}$ of the Faddeev term. If $f_\pi$ and $M$ are taken fixed as $\beta$ is varied this is indeed the case. This parametrization will be referred to as the \emph{fixed bound parametrization} since the energy satisfies the BPS inequality with an energy $E_{BPS}$ that is constant with $\beta$.

But if $M^2$ is allowed to vary with $\beta$, then there is no problem taking the $\beta=1$ limit. In particular, the \emph{Ward parametrization}~\cite{ward2004skyrmions},
\begin{align*}
\frac{f_\pi^2}{2}=\frac{1}{4\pi^2\left(3-\beta\right)},\qquad \frac{f_\pi^2}{2M^2}=\frac{1-\beta}{4\pi^2\left(3-2\beta\right)},
\end{align*}
is based on requiring that the identity map from a base space with spherical $S^3$ geometry to the $S^3$ target space has unit energy for all $\beta$, and it leads to a fairly constant dependence on $\beta$ of the energy of a $Q=1$ Skyrmion in flat space as well. No matter which parametrization for $f_\pi$ and $M$ is chosen, the results for any other parametrization may be recovered by adjusting the energy and length scales. Table \ref{table SqSkyrme} involves a simulation in the fixed bound parametrization, but the rescaled results agree with Ward up to an error of $\sim0.1\%$ from finite size effects.

\begin{table}
	\begin{center}
		\begin{tabular}{c|c|c|c|c}
			$\beta$ & $E/E_{BPS}$ & $E_{h}/E_{BPS}$ & $E_{W}$ &  $\frac{\langle (J_\mu^1)^2 \rangle}{\langle (J_\mu^3)^2 \rangle}$ \\
			\hline \hline
			0.0 &1.2323 &1.2331 & 1.2323& 1.0 \\
			0.1 &1.2339 &1.2348& 1.2324& 0.9873 \\
			0.2 &1.2392 &1.2403 &1.2324 & 0.9737 \\
			0.3 &1.2497 &1.2513 & 1.2322& 0.9592 \\
			0.4 &1.2679 &1.2702 & 1.2319& 0.9439 \\
			0.5 &1.2981 &1.3015 & 1.2315& 0.9279 \\
			0.6 &1.3486 &1.3535 & 1.2311& 0.9103 \\
			0.7 &1.4370 &1.4442 &1.2309 & 0.8912 \\
			0.8 &1.6111 &1.6224 & 1.2316& 0.8695 \\
			0.9 &2.0530 &2.0650 & 1.2269& 0.8519\\

		\end{tabular}
		\caption{A simulation of a $Q=1$ soliton in the squashed Skyrme model. $E$ is the energy in the fixed bound parametrization. The simulation was carried out on a cubic lattice with $100^3$ sites (except for $\beta=0.9$ where the length was doubled to $200^3$) and lattice spacing $a=0.2$ in units where $M=1$. An arrested Newton flow method was used for the minimization as described in~\cite{battye2002skyrmions}, with the time evolution implemented by a fourth-order Runge-Kutta method with time step $\Delta t = 0.1$. $E_h$ is the optimal energy in the spherically symmetric hedgehog ansatz for this same parametrization. $E_W$ is the energy in the Ward parametrization which was found by rescaling $E$. To better indicate the departure from the hedgehog ansatz, the values of $\left(J^1\right)^2$ and $\left(J^3\right)^2$ are averaged over the domain of the simulation and compared.}
		\label{table SqSkyrme}
	\end{center}
\end{table}

Any parametrization which allows for a well-defined $\beta=1$ limit will involve the energy $E_{BPS}$ in the BPS bound \eqref{eq BPS bound} tending to zero. This makes sense since in the Faddeev-Niemi model the minimal energy solutions obey a weaker $E\geq K Q^{3/4}$ inequality for some value of $K$~\cite{vakulenko1979stability,ward1999hopf} and moreover the minimal energy Hopfions found numerically~\cite{battye1998knots,battye1999solitons,sutcliffe2007knots} appear to come close to saturating this bound. For $\beta$ close to but less than $1$, the energies of solitons with small values of $Q$ may be very close to the energies in the Faddeev-Niemi model, and this is not disallowed by \eqref{eq BPS bound} since the value of $E_{BPS}$ may be very small. But no matter how small $E_{BPS}$ may be, eventually for large enough $Q$, $E_{BPS}Q> K Q^{3/4}$. So for $\beta<1$ the energies of the large $Q$ solitons can not scale asymptotically as $Q^{3/4}$, and thus if the Faddeev-Niemi model indeed has this asymptotic behavior there must be a dramatic difference for large $Q$ solitons if $\beta$ is even slightly below $1$.

\subsection{Position curves and baryon strings}\label{sec Position curves and strings}

Intuitively a Hopfion is often described as a loop of string whereas a single Skyrmion in the Skyrme model is spherically symmetric and multiple Skyrmions form polyhedral clusters. While we have shown that the baryon charge and Hopf charge are identical, let us comment a bit more on how these two pictures are resolved.

The $Q=1$ Skyrmion in the Skyrme model satisfies the hedgehog ansatz,
\begin{align}
	U(x^\mu) = \cos f(r)\,I+ i \sin f(r)\frac{x^i}{r}\sigma^i,\label{def Hedgehog}
\end{align}
for some radial profile function $f(r)$ which equals $\pi$ at $r=0$ and vanishes at infinity. Considering \eqref{def U and z} and \eqref{def S and z}, the third component of the unit vector $S$ field in the hedgehog ansatz is,
$$S^3=\cos^2 f -\frac{x^2+y^2-z^2}{r^2}\sin^2 f.$$
The boundary condition on the $S$ field at infinity $S^3=+1$ is also satisfied along the $z$ axis, and the furthest departure from the boundary condition $S^3=-1$ is satisfied in a loop in the $xy$-plane with radius $r_0$ such that $f(r_0)=\pi/2$. A curve such as this where $S^3=-1$ is referred to as the \emph{position curve}, and it may be thought of as the core of the Hopfion.

\begin{figure}[t]
\centering
\includegraphics[width=\textwidth]{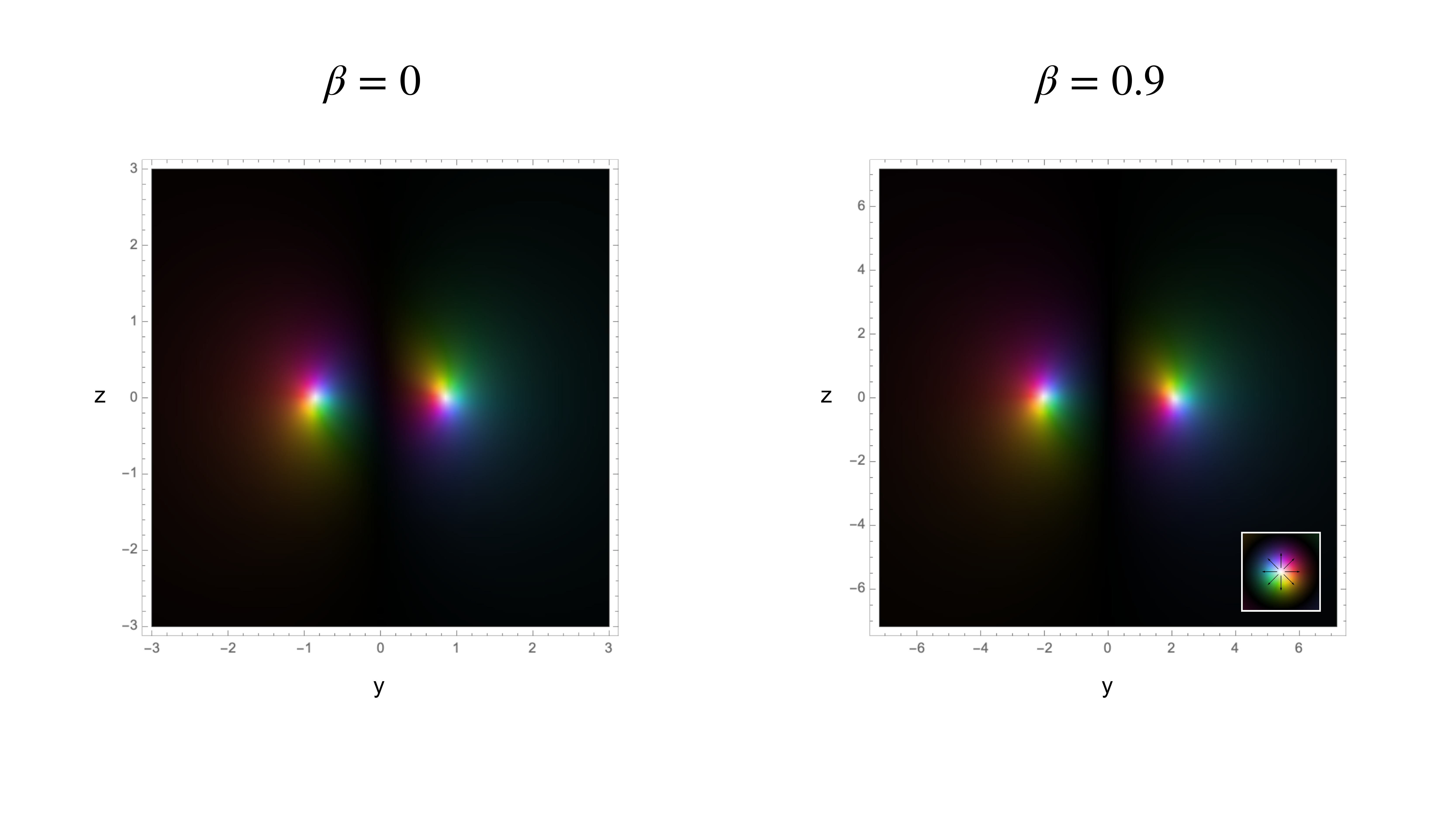}
\caption{On the left is a cross-section in the $yz$-plane of the hedgehog Skyrmion in the Skyrme model ($\beta=0$), mapped to the unit vector $S$ field, as discussed in the text. The
{color scheme follows the typical conventions in Lorentz transmission electron microscopy, where the}
hue of a color denotes the azimuthal angle of $S$, and the brightness denotes the polar angle. The limiting case of black represents the boundary condition at infinity, and white denotes the center of the position curve describing the core of a Hopfion. On the right is the minimal energy $Q=1$ soliton in the squashed Skyrme model at $\beta=0.9$ in the fixed bound parametrization. Both plots may be compared to similar plots for true Hopfions with gauge invariance such as Fig. 1b of~\cite{sutcliffe2017skyrmion}, and  Fig. 1b and Fig. 3 in~\cite{kent2021creation}.}
\label{fig Hedgehog cross section}
\end{figure}

A cross-section in the $yz$-plane of the Skyrmion at both $\beta=0$ and $\beta=0.9$ is plotted in Fig. \ref{fig Hedgehog cross section}. The two intersections of the position curve loop with the plane are clearly seen, and it may be seen from the colors representing the orientation of $S$ how the 2D magnetic Skyrmion charge in the $yz$-plane (quantified by $F_{23}$ \eqref{eq F tensor S}) is concentrated around the position curve. Note that due to the dependence on the $J^3$ field the energy density and baryon charge of the hedgehog Skyrmion are actually spherically symmetric and not concentrated near the position curve. But even the $Q=1$ Hopfion in the Faddeev-Niemi model at $\beta=1$ is approximately spherically symmetric in this sense as well, as was noted by Ward~\cite{ward2004skyrmions} and is seen by the extent to which the hedgehog ansatz fits the data in Table \ref{table SqSkyrme}.

\begin{figure}[t]
\centering
\includegraphics[width=0.8\textwidth]{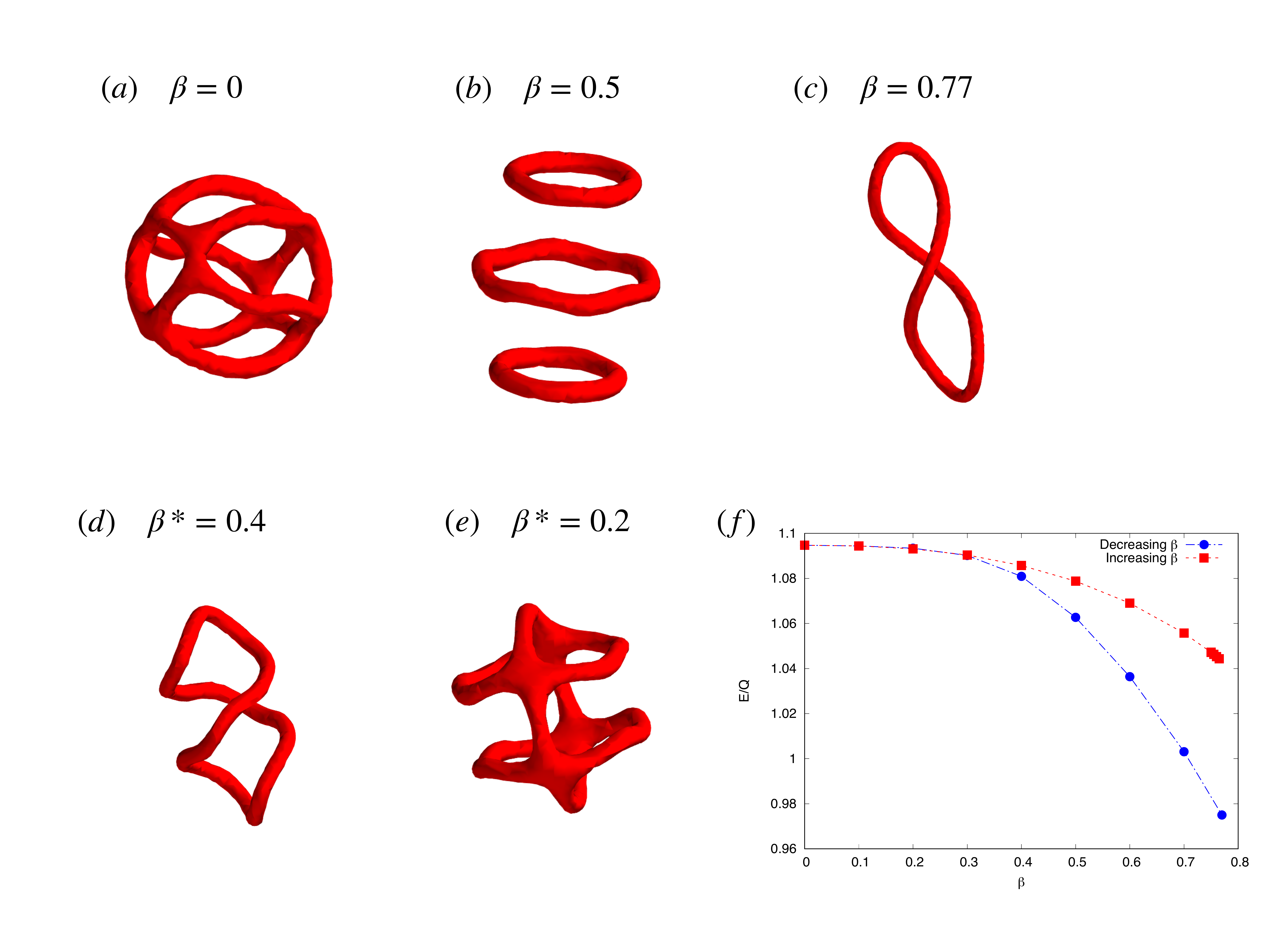}
\caption{In (a) a position curve of a $Q=7$ Skyrmion at $\beta=0$ is shown. For small $\beta>0$ this settles to a three loop configuration as in (b). This solution may be tracked for increasing $\beta$, but it develops an instability at $\beta=0.77$ and settles to a distinct buckled loop configuration (c). This buckled loop branch may be continued towards decreasing $\beta$ (indicated by an asterisk) in (d) and (e), and eventually it becomes degenerate in energy with the solution (a). The energy per charge in the Ward parametrization is plotted in (f), with red indicating the three loop branch shown in (b) and blue indicating the buckled loop branch of (c,d,e).}\label{fig Charge 7}
\end{figure}

The position curve may also be considered for higher charge solitons in the Skyrme model, such as the $Q=7$ case in Fig. \ref{fig Charge 7}. At $\beta=0$ the position curve self-intersects much like the $\chi$ solutions found by Sutcliffe in the Faddeev-Niemi model~\cite{sutcliffe2007knots}. {At $\beta=0.2$ the position curve transforms to three loops and we were originally anticipating that this solution would smoothly deform to a trefoil knot as was seen at $\beta=1$ in the Faddeev-Niemi model~\cite{battye1998knots,battye1999solitons}. However it appears that there are actually many locally stable solution branches which may exchange roles as the global minimum as the deforming parameter is adjusted.
	
	In particular, we found that, at $\beta = 0.77$, a three loop configuration becomes unstable and evolves to a buckled loop with lower energy. We explored this new configuration by decreasing $\beta$ all the way back to $\beta=0$, in which case the buckled loop transforms to a self-intersecting solution which is degenerate with our original solution in Fig. \ref{fig Charge 7}(a).
	This wealth of local minima will be seen again in the results of the lattice model in \ref{sec charge-10}, and has also been seen in systems such as the Skyrme model with a non-zero pion mass \cite{bjarke2022smorgaasbord}.}

Note that at $\beta=0$ the global symmetry of system under right multiplication is increased from $U(1)$ to $SU(2)$. Any soliton configuration $U$ may be transformed to a new field $U \rightarrow V^{-1}U V$ with the same boundary conditions at infinity and the same energy. For a general $V\in SU(2)$ this transformation will not leave the position curve invariant. For the hedgehog configuration this ends up being equivalent to the degeneracy of the solution under spatial rotations, but for higher charge configurations the shape of the position curve itself may change. However for $\beta>0$ the symmetry is reduced to a {$U(1)\rtimes Z_2$} subgroup\footnote{ {The extra discrete $Z_2$ global symmetry arises from those internal $SU(2)_R$ transformations which flip the third axis in isospin space. Concretely the $Z_2$ subgroup may be chosen as $\{ I,\sigma^1\}\subset SU(2)_R$.}} which leaves the locus of the position curve unchanged, and only translates the field along the position curve.

Some insight may be gained by considering the structure of the $U$ field around the position curve in the hedgehog ansatz in \eqref{def Hedgehog} and Fig. \ref{fig Hedgehog cross section}, and abstracting this to a new ansatz of a cyllindrically symmetric straight string with a position curve aligned with the $z$-axis,
\begin{align}
	U(\rho,z,\phi)= \cos g(\rho) e^{-i\phi\sigma^3}+i\sin g(\rho)\left(\cos\left(\frac{2\pi z}{L}\right)\sigma^1+\sin\left(\frac{2\pi z}{L}\right)\sigma^2\right).
\end{align}
Here $g(\rho)$ is some new profile function depending on $\rho=\sqrt{x^2+y^2}$, and $L$ is some parameter describing the rate of twisting along the string. The profile function vanishes at infinity and $g(0)=\pi/2$. The baryon charge \eqref{def Q Skyrme} integrated over a length $\Delta z$ is found to be,
\begin{align*}
	Q= -\frac{1}{L}\int dz d\rho \,g^\prime(\rho)\sin \left(2g(\rho)\right) =\frac{\Delta z}{L},
\end{align*}
so every segment of length $L$ has baryon charge $1$.

Outside the core of the string, where $g\approx 0$, $U$ is restricted to a $U(1)$ subgroup, and the principal chiral model effectively reduces to a 3D XY model. Unless the $U(1)$ subgroup is gauged (as it is at $\beta=1$) the energy per length of an isolated straight string will be logarithmically divergent.

To better understand the energy per length due to the core of the string, the structures involved in the energy density \eqref{lagr Ward} may be expressed in terms of the ansatz,
\begin{gather}
	(J^1)^2+(J^2)^2=g^{\prime2}+\left[\left(\frac{2\pi}{L}\right)^2+\frac{1}{\rho^2}\right]\cos^2g \sin^2 g \non
(J^3)^2=\left(\frac{2\pi}{L}\right)^2 \sin^4g + \frac{1}{\rho^2}\cos^4g	\non
	\frac{1}{8}\left(F_{\mu\nu}\right)^2=g^{\prime 2}\left[\left(\frac{2\pi}{L}\right)^2+\frac{1}{\rho^2}\right]\cos^2g \sin^2 g\non
-\frac{1}{4}\text{Tr}[J_\mu,J_\nu]^2=\left[\sin^2g \left(\frac{2\pi}{L}\right)^2 +\cos^2g \frac{1}{\rho^2}\right]g^{\prime 2}+\left(\frac{2\pi}{L}\right)^2\frac{1}{\rho^2}\cos^2g \sin^2g.\label{eq Energy string ansatz}
\end{gather}
The $\rho^{-2}$ term in the $\left(J^3\right)^2$ structure is what causes the infrared divergence of an isolated string. Of course this is typical for strings or vortices and does not preclude either a network of oppositely oriented long strings or strings forming closed loops with radius of curvature much larger than the string thickness, in which cases this ansatz may still be useful. In the latter case, due to the reduction of the model to a global $U(1)$ theory outside the string core, and due to the expression \eqref{def F tensor} relating the curl of $J^3$ to the dual of $F$ tensor which has a constant $2\pi$ flux across the string core, the $J^3$ field may be calculated outside the string core using the Biot-Savart formula, much as is done calculating the fluid velocity outside knotted loops of vorticity~\cite{kleckner2013creation}.

The energy inside the core may be calculated by optimizing the above expressions for the energy \eqref{lagr Ward} and \eqref{eq Energy string ansatz} over profile functions $g$ and length per baryon charge $L$. An important special case is that of Faddeev-Niemi model itself at $\beta=1$, in which case the energy contribution from the $J^3$ field outside the core vanishes. To compare with previous results, we may use the parametrization in Battye and Sutcliffe~\cite{battye1999solitons} where $f^2_\pi/2=4$ and  $M\sqrt{1-\beta}\rightarrow 1/2$. Then the optimum energy of the straight string ansatz is found to be $E \approx 396$ per length $L\approx 3.95$. The energy per charge already agrees reasonably well with the unstable toroidal solutions of Battye and Sutcliffe in Table 1 and Figure 9 of~\cite{battye1999solitons}, which may be expected to become closer to the straight Skyrmion string ansatz as the charge increases.

\section{A toy model of a frustrated magnet}\label{sec 3}
Now we will introduce a simple spin system which at lowest order in the continuum approximation reduces to the same squashed sphere non-linear sigma model discussed in the previous section. The higher order derivative terms which may stabilize topological defects will be different from the rotationally symmetric Skyrme and Faddeev-Niemi terms, but are in many respects qualitatively similar.

\subsection{A description of the lattice model} \label{sec Lattice model intro}

The system is defined on an ordinary cubic lattice with lattice spacing $a$, and each site $x$ has three real unit vector spins $S^i_r(x)$, where the $i$ index refers to the three components of the unit vector, and the $r$ index labels the distinct spins at the site. The dot product between any two spins at a given site is constrained to be equal to a parameter $\kappa$ which is fixed for the entire system, i.e. $S_r(x)\cdot S_s(x)=\kappa$ for $r\neq s$ and all $x$. So the three spins at each site act like a rigid body with an orientation which may be described by a matrix $R(x)\in SO(3)$.
The spins may be written in terms of this $R(x)$ and a basis $e_r$ which does not depend on $x$,
\begin{gather}
S^i_r(x)=R(x)^i_{\,j}e^{j}_r,\qquad
e^j_{r}\equiv
\left(\begin{array}{ccc}
\frac{\sqrt{3}}{2}\sin\theta& 0 &-\frac{\sqrt{3}}{2}\sin\theta\\
-\frac{1}{2}\sin\theta &\sin\theta&-\frac{1}{2}\sin\theta\\
\cos\theta & 	\cos\theta & 	\cos\theta
\end{array}\right),\label{def R e}
\end{gather}
where the three vectors $e_r$ are represented as a matrix with $r$ referring to different columns. The fixed parameter $\theta$ in this basis is directly related to the parameter $\kappa$,
\begin{align}
 S_r(x)\cdot S_s(x)=\kappa \equiv \frac{3}{2}\cos^2\theta-\frac{1}{2}.\label{def Kappa theta}
\end{align}

The spins interact as a frustrated classical Heisenberg model, with a ferromagnetic coupling $K_1<0$ between nearest neighbors at a distance of one lattice spacing $a$, and an antiferromagnetic coupling $K_2>0$ between sites at a distance of $2a$, which are indicated by doubled angled brackets in a slight abuse of notation,\footnote{For simplicity in this toy model, sites at the nearer distances of $\sqrt{2}a$ and $\sqrt{3}a$ are not taken to interact.}
\begin{align}
H&=K_1\sum_{r,\,\langle x,y\rangle}S_r(x) \cdot S_r(y)+K_2\sum_{r,\,\llangle x,y\rrangle}S_r(x) \cdot S_r(y).\label{ham S}
\end{align}
Note that a given spin $S_r(x)$ only interacts with spins $S_r(y)$ with the same `species' $r$.

This model {was originally inspired} by the treatment of spins interacting on a pyrochlore lattice in~\cite{BatistaEtAl2018}, where in that case $r$ takes four values corresponding to the four sites of the tetrahedral cells of the pyrochlore lattice. The dot product between spins $\kappa$ in that case is fixed so that the spins are in an `all-in-all-out' configuration which is preferred {in the presence of biquadratic spin interactions}. If we formally allow $\kappa$  to be a tunable parameter and restrict the interaction to third-nearest-neighbor sites so only spins with the same value of $r$ interact we obtain a very similar model to that considered here.

%This model can be regarded as the low-energy description for the broad class of realistic 3D noncollinear magnets, where SO(3) rotation is the only low-energy mode while other modes are gapped out. For instance, the Heisenberg model on pyrochlore lattice was demonstrated to have gapless SO(3) mode, where other modes are gapped in the presence of biquadratic spin interactions [34]. The parameter κ is expected to vary for different 3D lattices and different magnetic anisotropies. As a result, the model study in this paper is expected to provide a unified qualitative description for a series of realistic materials that are described by different values of κ.

{The motivation for making these abstractions was to create a simple lattice model that still captures the main qualitative features of a broad class of realistic 3D noncollinear magnets which involve $SO(3)$ Goldstone modes. The `squashing' parameter $\kappa$ that measures the degree of collinearity} is expected to vary for different 3D lattices and different magnetic anisotropies. So the model studied in this paper is expected to provide a unified qualitative description for a series of realistic materials that are described by different values of $\kappa$.

Also note that at the limiting value $\kappa=1$ where the spins $S_r$ {are perfectly collinear}, the lattice model reduces to a 3D version of inversion-symmetric frustrated magnets
which have been previously considered in 2D as a host to magnetic Skyrmions~\cite{leonov2015multiply,lin2016ginzburg}. A 3D extension of these frustrated magnets has already been considered {in the collinear case} \cite{sutcliffe2017skyrmion}, and Hopfions were investigated and the analogy to the Faddeev-Niemi model was pointed out.

But in the opposite limit of $\kappa=0$ this model will instead be shown to be closely analogous to the Skyrme model, so this toy model bears the same relationship to the effective theory of frustrated magnets in~\cite{sutcliffe2017skyrmion} as the squashed Skyrme model~\cite{nasir2002effective,ward2004skyrmions} discussed in the previous section bears to the Faddeev-Niemi model.

\subsection{Effective theory in the continuum limit}\label{sec Lattice model effective theory}

To show that this analogy is valid, let us now turn to the effective continuum description of the model. Following a similar procedure to Dombre and Read's continuum description of the triangular antiferromagnet~\cite{dombre1989nonlinear}, the Hamiltonian can be described up to fourth order in derivatives in terms of continuous fields $S^i_r(x)$,
\begin{align}
H=-\frac{1}{2a}\left(K_1+4K_2\right)\sum_\mu \int d^3 x \left(\partial_\mu S_r\right)^2 +\frac{a}{24}\left(K_1+16 K_2\right)\sum_\mu \int d^3 x \left(\partial^2_\mu S_r\right)^2. \label{ham Continuum}
\end{align}
Now since rotational symmetry is broken by the fourth order terms, any sums over the spatial index $\mu$ will always be indicated explicitly, although sums over internal indices like $r$ or $i$ are still implied by the summation convention or context. The lack of rotational symmetry in the fourth order terms is the main difference between the effective description of this toy model and that considered by Lin and Hayami~\cite{lin2016ginzburg}. Here the interaction between the neighbors at distances $\sqrt{2}a$ and $\sqrt{3}a$ was set to zero whereas in~\cite{lin2016ginzburg,sutcliffe2017skyrmion} it was implicitly tuned to maintain rotational symmetry in the fourth order terms. Note that in the absence of any tuning such cubic anisotropies would generically be expected to be present, {and the presence of isotropy in these higher derivative terms is not essential for the stabilization of topological defects.}

For this Hamiltonian to have stable topological defects it is easily shown by an argument along the lines of Derrick's theorem~\cite{derrickstheorem} that the coefficients of both the second and fourth order terms must be positive,
$$-K_1 > 4 K_2 > -\frac{1}{4}K_1.$$

Moreover, for the Skyrmion size to be much larger than the lattice spacing and this continuum description to be valid we must be close to the Lifshitz transition $K_2= -\frac{1}{4}K_1$ where the sign of the quadratic term changes from positive to negative. Suppose that a Skyrmion field configuration has some length scale $L$ representing the radius, and the parameters are displaced from the Lifshitz transition by some small positive quantity $\epsilon$,
\begin{align}
K_2= -\left(\frac{1}{4}-\epsilon\right)K_1.\label{def epsilon}
\end{align}
Then it can be shown that radius of the Skyrmion is on the order
$L\sim \epsilon^{-1/2}a,$ where the exact coefficient depends on dimensionless integrals over the field configuration.

Now to proceed and better illustrate the connection to the squashed sphere sigma model in Sec.~\ref{sec Squashed sphere}, the spins $S_r$ may be written in terms of the rotation matrix field $R(x)$ using \eqref{def R e}. The quadratic terms become
$$\sum_\mu \int d^3 x \left(\partial_\mu S_r\right)^2 = \sum_\mu \int d^3 x \text{Tr}\left[\partial_\mu R^{-1}\partial_\mu R \,e_r\otimes e_r\right],$$
where
\begin{align}
	e_r\otimes e_r = \text{diag}\left(1-\kappa,\,1-\kappa,\, 1+2\kappa\right).
\end{align}
For $\kappa=0$ this is clearly equivalent to the principal chiral model \eqref{lagr PCM U}, except that it is expressed in terms of $R\in SO(3)$ rather than $U\in SU(2)$. For $\kappa \neq 0$, the components of the diagonal matrix $e_r\otimes e_r$ will take different values and this will become a squashed sphere model. This can be seen by expressing the model in terms of the currents $J$ \eqref{def J currents}, which may also be expressed in terms of the $SO(3)$ matrix,
\begin{align}
\left(R^{-1}\partial_\mu R\right)_{ij} = 2\epsilon_{ijk}J^k_\mu.\label{eq J R}
\end{align}
Using this identity, the quadratic terms become
\begin{align}
\sum_\mu \int d^3 x \left(\partial_\mu S_r\right)^2 = 4(\kappa+2)\sum_\mu \int d^3 x \left[\left(J^i_\mu\right)^2-\frac{\kappa}{\frac{1}{3}\left(\kappa+2\right)}\left(J^3_\mu\right)^2\right].\label{lagr Quadr}
\end{align}
This is precisely the squashed sphere model in \eqref{lagr Squashed J}, with the parameter $\beta$ expressed in terms of $\kappa$. The overall dimensionfull parameter $f_\pi$ in the squashed sphere model depends on the prefactor of the quadratic terms given in the full Hamiltonian \eqref{ham Continuum}, and it is seen to be on the order $f_\pi \sim \left(\epsilon|K_1|a^{-1} \right)^{1/2}$.

Exactly the same chain of steps may now be followed to express the quartic terms of the Hamiltonian in terms of the $J$ fields and the parameter $\kappa$. After some calculation,
\begin{align}
	\sum_\mu \left(\partial^2_\mu S_r\right)^2 = &8(1-\kappa)\sum_\mu \left[\left(\partial_\mu J_\mu^i\right)^2+4\left(J_\mu^i J_\mu^i\right)^2\right]+ 12\kappa \sum_\mu \left[\left(\mathcal{D}_\mu J_\mu^a\right)^2+4\left(J_\mu^a J_\mu^a\right)^2\right],\label{lagr Quartic}
\end{align}
where as discussed previously, $i$ runs over all components $1, 2, 3$, and $a$ is only taken over $1, 2$. The covariant derivative with respect to the gauge symmetry defined in \eqref{eq J gauge transf} is
$$\mathcal{D} J^a\equiv \partial J^a + 2\epsilon_{ab3}J^3\,J^b.$$
Note that the continuum model is completely gauge symmetric at $\kappa=1$, which must be the case considering that in the lattice model all three spins at each site are pointing in the same direction, so the rotation field $R(x)$ is only fixed up to rotations about the spin axis.

This continuum description of the model in equations \eqref{ham Continuum}\eqref{lagr Quadr}\eqref{lagr Quartic} will later be applied to calculate the energies of highly symmetric ansatzes for Skyrmion configurations, and the results will be compared with Skyrmions found in a numerical simulation of the lattice model described below.

\subsection{Numerical simulation and unit charge Skyrmions}\label{sec Lattice model numerical results}

In this section, we perform numerical simulations directly on the lattice model~\eqref{ham S}, which captures the higher order terms and spatial anisotropies that we neglected in the continuum model~\eqref{ham Continuum}. Strictly speaking, the energy barriers between different topological sectors are no longer infinity on a discrete lattice. Consequently, when the Skyrmion size is not significantly larger than the lattice spacing $a$, it could be unstable towards tunneling into the vacuum state. In such situations, it is beneficial to fully relax the assumed Skyrmion configuration and check the stability.

Typically, the local minima of the classical {spin} models can be found by the Landau-Lifshitz-Gilbert dynamics~\cite{LandauLifshitz1992,Gilbert2004} that works directly on the spin degrees of freedom. To enforce the constraint $S_r(x)\cdot S_s(x)=\kappa$, a penalty term can be included in the model, which slightly complicates the computation.

To avoid such complication, we work directly with the rotation matrix $R(x)$. The spin-spin interaction between site $x$ and $y$ can be written as
\begin{equation}
\sum_r S_r(x) \cdot S_r(y) =  \text{Tr} \left[
	R^{-1}(x)R(y) e_r\otimes e_r
	\right]. \label{eq:spin-spin_rotation}
\end{equation}

There are a few representations that can be used for the rotation matrix. To avoid ``Gimbal lock'', we use the quaternion representation in this work:
\begin{equation}
	R=\begin{pmatrix}q_{0}^{2}+q_{1}^{2}-q_{2}^{2}-q_{3}^{2} & 2q_{1}q_{2}-2q_{0}q_{3} & 2q_{1}q_{3}+2q_{0}q_{2}\\
		2q_{1}q_{2}+2q_{0}q_{3} & q_{0}^{2}-q_{1}^{2}+q_{2}^{2}-q_{3}^{2} & 2q_{2}q_{3}-2q_{0}q_{1}\\
		2q_{1}q_{3}-2q_{0}q_{2} & 2q_{2}q_{3}+2q_{0}q_{1} & q_{0}^{2}-q_{1}^{2}-q_{2}^{2}+q_{3}^{2}
	\end{pmatrix},
\end{equation}
where the quaternion $\bm{q}=(q_0,q_1,q_2,q_3)^T$ is a four-vector satisfying
\begin{equation}
	q_0^2+q_1^2+q_2^2+q_3^2=1.
\end{equation} Note that the quaternion $\bm{q}$ is related to the complex numbers $z^0$ and $z^1$ introduced in Eq.~\eqref{def U and z} by
\begin{equation}
	z^0 = -q_2 - i q_1,\quad z^1=q_0 + i q_3.
\end{equation}

The Hamiltonian~\eqref{ham S} is now expressed in terms of the quaternions:
\begin{equation}
	\begin{split}
		H &= 3 K_1 \sum_{\langle x,y \rangle} \left\{ q_0^2(x,y) - \cos^2 \theta \left[ q_1^2(x,y) + q_2^2(x,y) \right] + \cos (2\theta) q_3^2(x,y) \right\} \\
		&\quad + 3 K_2 \sum_{\llangle x,y \rrangle} \left\{ q_0^2(x,y) - \cos^2 \theta \left[ q_1^2(x,y) + q_2^2(x,y) \right] + \cos (2\theta) q_3^2(x,y) \right\},
	\end{split}\label{ham quaternion}
\end{equation}
where the quaternion product is defined as
\begin{equation}
\bm{q}(x,y)\equiv \overline{\bm{q}(x)} \bm{q}(y)
=
\begin{pmatrix}
\quad q_{0}(x)q_{0}(y)+q_{1}(x)q_{1}(y)+q_{2}(x)q_{2}(y)+q_{3}(x)q_{3}(y)\\
-q_{1}(x)q_{0}(y)+q_{0}(x)q_{1}(y)-q_{2}(x)q_{3}(y)+q_{3}(x)q_{2}(y)\\
-q_{2}(x)q_{0}(y)+q_{0}(x)q_{2}(y)-q_{3}(x)q_{1}(y)+q_{1}(x)q_{3}(y)\\
-q_{3}(x)q_{0}(y)+q_{0}(x)q_{3}(y)-q_{1}(x)q_{2}(y)+q_{2}(x)q_{1}(y)
\end{pmatrix},\label{quaternion prod}
\end{equation}
{and the quaternion conjugate is $\overline{\bm{q}}\equiv (q_0,-q_1,-q_2,-q_3)^T$}.

The Skyrmion solutions are local minima of the lattice model. Consequently, they can be obtained by local minimization algorithms from initial spin configurations not too far away from the minima. In this work, we use the low-storage BFGS method~\cite{nlopt, Nocedal1980, Liu1989} for the minimization. When the spin configurations get close enough to the minima, we switch to the overdamped Langevin dynamics to avoid being trapped in saddle points:
\begin{equation}
\frac{d \bm{q}(x)}{dt} = \bm{f}(x) - \left[ \bm{f}(x) \cdot \bm{q}(x)\right] \bm{q}(x), \label{langevin}
\end{equation}
where the force $\bm{f}(x)$ is defined as
\begin{equation}
\bm{f}(x)= -\frac{dH}{d \bm{q}(x)}.
\end{equation}
The overdamped Langevin dynamics~\eqref{langevin} is integrated by the explicit fourth order Adams-Bashforth method as predictor and the implicit fourth order Adams-Moulton method as corrector. The time step $dt$ is chosen as $dt=0.01/|K_1|$ for $\kappa=0$ and $0.625$, and $dt=0.005 /|K_1|$ for $\kappa \approx 0.955\, (\theta=\pi/18)$.

Before discussing the numerical solutions of Skyrmions, we discuss the energy of the vacuum here. For the ferromagnetic state $\bm{q}(x)=\bm{q}(y)$, the quaternion product~\eqref{quaternion prod} is $\bm{q}(x,y)=(1,0,0,0)^T$. Consequently, the energy is
\begin{equation}
E_{\text{FM}}=
\begin{cases}
9N\left(K_{1}+K_{2}\right), & \text{periodic boundary condition (PBC)}\\
9N\left(K_{1}+K_{2}\right)+9L^{2}(K_{1}+2K_{2}), & \text{fixed boundary condition (FBC)}
\end{cases}
\end{equation}
where $N=L^3$ is the total number of lattice sites. In this paper, we fix $L=128$. Here, the two types of boundary conditions {(BC)} differ in how a quaternion $\bm{q}(x)$ with $x$ inside the boundary is connected to another quaternion $\bm{q}(x+\delta x)$ with $x+\delta x$ outside the boundary: in PBC $x+\delta x$ is translated back to the lattice by displacement vector $(mL,nL,lL)$ where $\{m,n,l\}$ are integers; in FBC we simply set $\bm{q}(x+\delta x) = (1,0,0,0)^T$. {In the following text, we always define the energy $E$ as the total energy where $E_\text{FM}$ has been subtracted.}

\begin{figure}[tbp]
\centering
\includegraphics[width=1\textwidth]{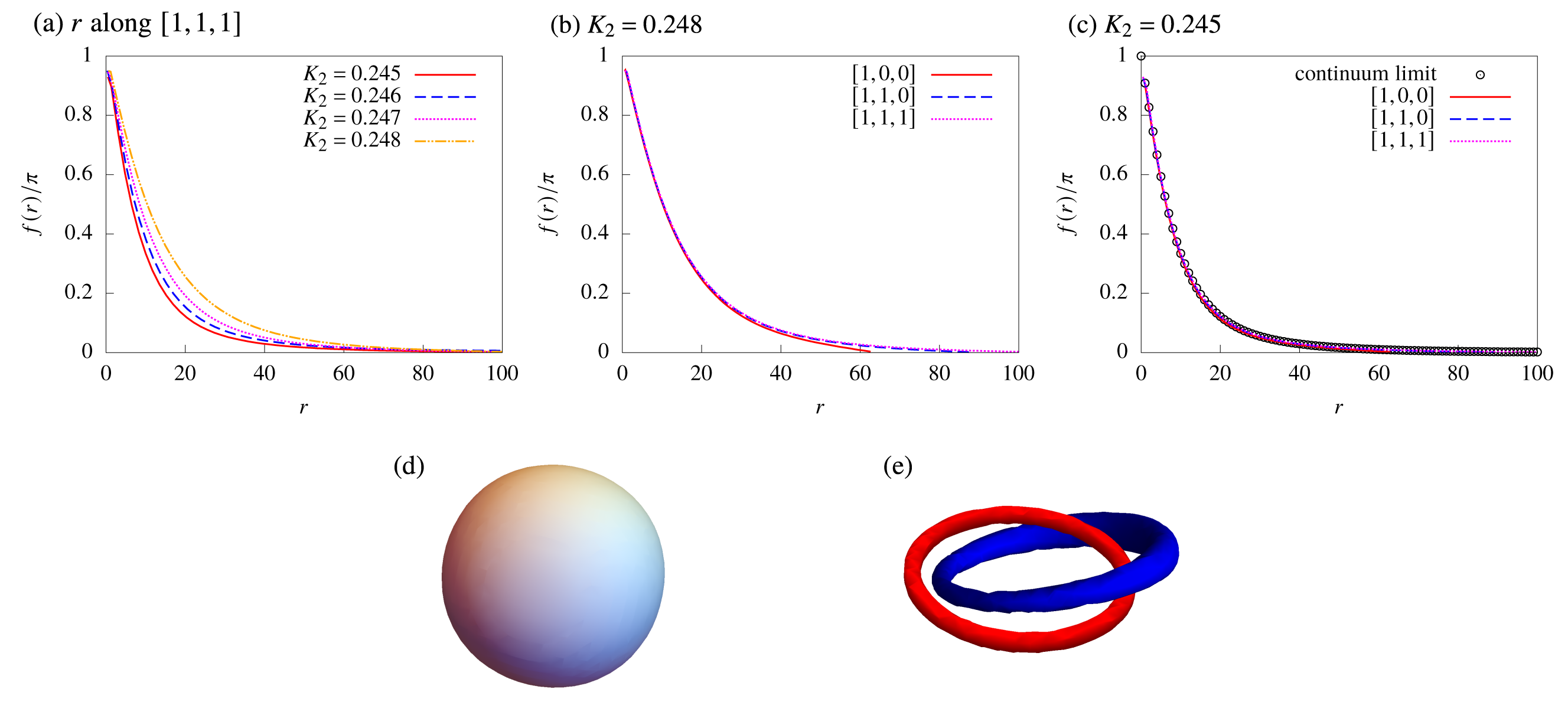}
\caption{The relaxed unit charge Skyrmion solutions on a $128\times 128\times 128$ lattice with $K_1=-1$, $\kappa=0$, and PBC. (a) The profile functions for different choices of $K_2$, where [1,1,1] is chosen as the radial direction. We have also computed the curves for $K_2=0.248$ with FPC, and the profile functions appear to be the same as the PBC ones by eye. (b) The profile functions for $K_2=0.248$ along different high-symmetry directions. (c) The profile functions for $K_2=0.245$ along different high-symmetry directions, along with the solution of the continuum theory Eq.~\eqref{ham Continuum}. (d) The topological charge density isosurface with $K_2=0.248$. (e) The position curve (red) and the linking curve (blue).}
\label{fig profile lattice charge1}
\end{figure}

Now we move to the discussion of the unit charge Skyrmion on the lattice. Figure~\ref{fig profile lattice charge1} shows the relaxed unit charge solutions of the lattice model at $\kappa = 0$, where local minima are obtained by the combination of minimization and Langevin dynamics. The Hopf charge $Q=1$ may be  immediately read out from the linking of the two curves in Fig.~\ref{fig profile lattice charge1}(e). The red curve is the \emph{position curve} which was discussed in Sec.\,\ref{sec Position curves and strings}, and is defined as the curve where the spin $S\equiv R\hat{z}$ takes the value $(0,0,-1)$. The blue curve will be referred to as the \emph{linking curve} and is instead where $S$ takes the value  $(0,-1/\sqrt{2},-1/\sqrt{2})$.

The linking of the position curve and linking curve gives a clear definition of the Hopf charge on a lattice, but there is an alternative method for defining the topological charge from a finite difference approximation to \eqref{def Q Skyrme}, where the $J$ fields are expressed in terms of the quaternion components. The latter method does not produce exact integer values for $Q$ {numerically}, but the departure from an integer value may be used as a rough estimate of the intrinsic `discretization error' that may be expected from using a continuum Hamiltonian \eqref{ham Continuum} in place of the exact lattice Hamiltonian \eqref{ham S}. {Indeed, the departure from the exact integer is found to be more significant when $K_2$ moves away from the Lifshitz point ($K_2/|K_1|=1/4$) causing a reduction of the Skyrmion size relative to the lattice constant.}

\begin{table}[tbp]
\centering
\begin{tabular}{|c|c|c|c|c|c|}
\hline
\diagbox[width=3.5cm]{$\qquad\kappa$}{$K_2$} & 0.245 (PBC) & 0.246 (PBC) & 0.247 (PBC) & 0.248 (PBC) & 0.248 (FBC) \\
\hline
0  & 91.713 & 82.670 & 72.184 & 59.530 & 59.734 \\
\hline
0.625  & \slashbox{}{} & \slashbox{}{} & \slashbox{}{} & {58.746} & {58.930} \\ \hline
0.955   & \slashbox{}{} & \slashbox{}{} & \slashbox{}{} & {57.739} & {57.896} \\ \hline
\end{tabular}
\caption{The relaxed energy {$E$} of the {unit charge} Skyrmion on a $128\times 128\times 128$ lattice with $K_1=-1$. The boundary conditions are indicated in the parentheses.}\label{table:E_Q1}
\end{table}

\begin{table}[tbp]
\centering
\begin{tabular}{|c|c|c|}
\hline
\diagbox[width=2.5cm]{$\qquad\kappa$}{BC} &  PBC & FBC \\
\hline
0 & 1.0  & 1.0  \\
\hline
0.625 & \quad 0.885 \quad  & \quad  0.886 \quad  \\ \hline
0.955 & \quad  0.834 \quad  & \quad  0.837 \quad  \\ \hline
\end{tabular}
\caption{ { $\langle \left( J_y^1\right)^2 \rangle/ \langle  \left( J_y^3\right)^2 \rangle$ of the unit charge Skyrmion on a $128\times 128\times 128$ lattice with $K_1=-1$ and $K_2=0.248$.}}\label{table:J_Q1}
\end{table}

%NOTES: These paragraphs are about how well the unit charge solutions agree with the hedgehog ansatz, this is why I was talking about the continuum computation. The profile function is defined through that ansatz in the continuum model. We exchanged emails in November where we agreed that detailed discussion of the approximation to the profile function in the lattice model was confusing here, and instead I paraphrased the discussion of this by "extracting rotation angles along various directions in the lattice."
% In any case it seems you think my discussion of the continuum model is distracting and I think the same about the details of profile functions found from the lattice model, so that's why I think it's best to just have a smaller paragraph.

These unit charge solutions at $\kappa=0$ are actually well described by the continuum hedgehog ansatz \eqref{def Hedgehog}, and their approximate rotational symmetry may be seen in the topological charge density isosurface plotted in Fig.~\ref{fig profile lattice charge1}(d).  The profile function may be found directly by minimizing the energy functional in the continuum theory, as discussed further in the next section. As shown in Fig.~\ref{fig profile lattice charge1}(c), the continuum profile function at $K_2=0.245$ agrees very closely with the profile function found by extracting the rotation angles along various directions in the lattice simulation. As $K_2$ is increased towards the Lifshitz point the Skyrmion size increases, as may be seen from the profile functions in Fig.~\ref{fig profile lattice charge1}(a). As shown in Fig.~\ref{fig profile lattice charge1}(b), at $K_2=0.248$ the Skyrmion size is comparable to the box size, and cubic anisotropies from the boundary conditions lead to the profile function being slightly different when calculated along three different high-symmetry directions on the cubic lattice.

{The parameter $\kappa$ (or equivalently $\theta$) allows us to interpolate between the $\text{SO}(3)$ and the $S^2$ target spaces. The energy $E$ of the relaxed unit charge Skyrmion for $\theta = \pi/6$ ($\kappa=0.625$) and $\theta = \pi/18$ ($\kappa \approx 0.955$) are recorded in Table~\ref{table:E_Q1}, which deviate only slightly from the $\kappa=0$ case. Small distortion of the topological charge isosurface also appears for $\kappa \neq 0$. To better indicate the departure from the $\kappa = 0$ limit, we also compare the average of $\left( J_\mu^1\right)^2$ to $\left( J_\mu^3\right)^2$ (see Table~\ref{table:J_Q1} for $\mu=y$). The results turn out to be quite similar to the ones shown in Table~\ref{table SqSkyrme} for the squashed Skyrme model.  }

\subsection{Higher charge Skyrmions and rational maps} \label{sec Rational map}

\begin{figure}[tbp]
\centering
\includegraphics[width=0.8\textwidth]{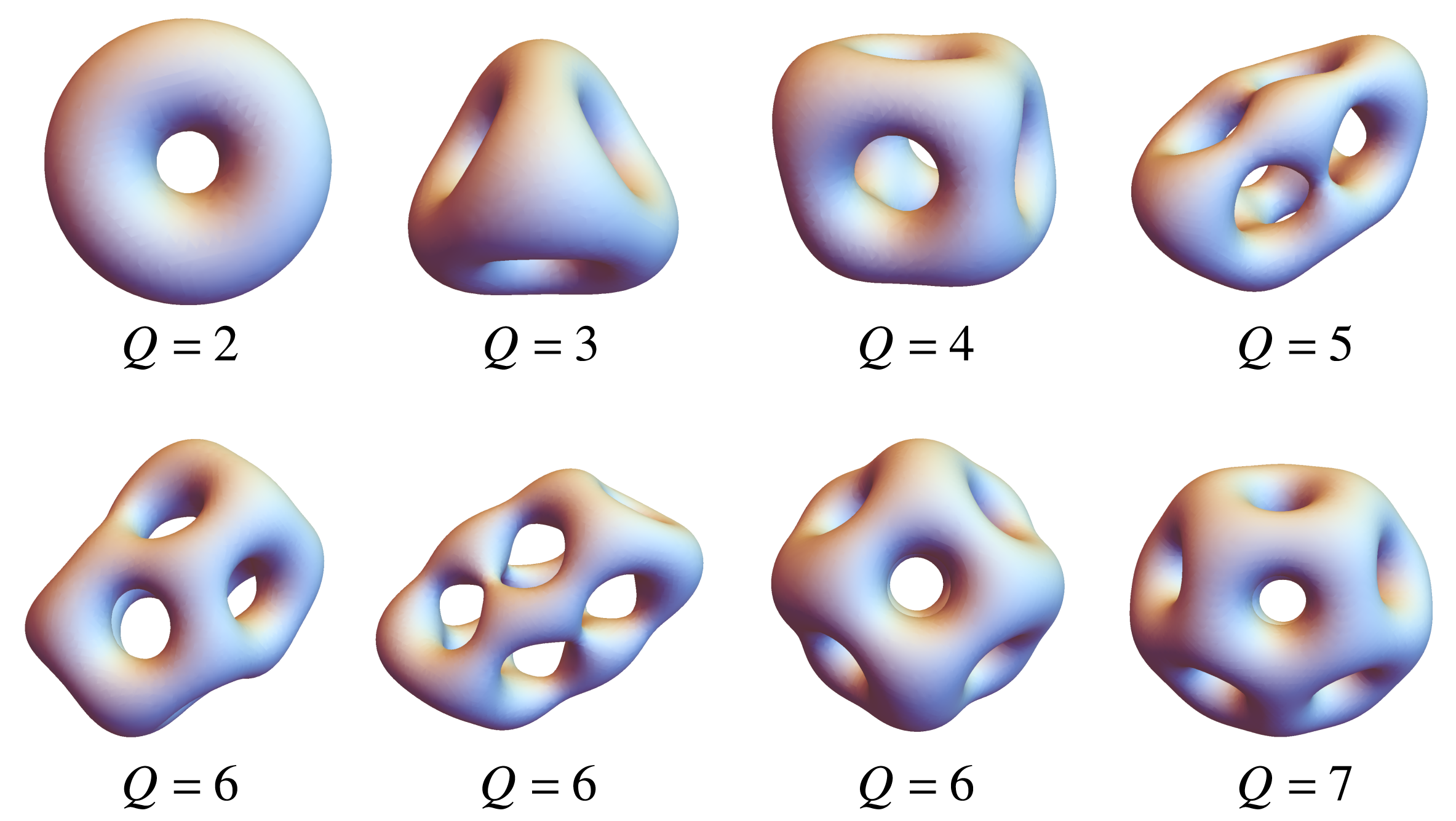}
\caption{Charge density isosurfaces of the relaxed higher charge Skyrmion solutions on a $128\times 128\times 128$ lattice with $K_1=-1$, $K_2=0.248$, and $\kappa=0$. The shapes for PBC and FBC are found to be the same by eye. Note for $Q=6$: all three isosurfaces are found to be stable for PBC, while only the first two on the left are found to be stable for FBC. }
\label{fig high charge}
\end{figure}

For Skyrmions with relatively low charge, we can create them by the method of ``merging''. For $Q=\{2,3,4\}$, we follow Ref.~\cite{MantonBook} by putting multiple $Q=1$ Skyrmions in the attractive channel and wait until the energy is fully minimized. For higher charge, to avoid missing the lowest energy solution, we use multiple ways of merging. In particular, for $Q=5$, we try two possible combinations: $Q=1+4$ and $Q=2+3$, which are found to relax to the same state; For $Q=6$, we try $Q=1+5$, $Q=2+4$, and $Q=3+3$, where two solutions are found; For $Q=7$, we try $Q=1+6$, $Q=2+5$ and $Q=3+4$, which all relax to the same solution. The energies of the solutions can be found in Table~\ref{table multi-charge}, and the charge density isosurfaces can be found in Fig.~\ref{fig high charge}.

\begin{table}[tbp]
\centering
\begin{tabular}{|c|c|c|c|c|c|c|c|c|c|}
\hline
\diagbox[width=2.0cm]{BC}{Q} & 1 & 2 & 3 & 4 & 5 & \multicolumn{3}{c|}{6} & 7\\
\hline
PBC & 59.530 & 56.274 & 53.883 & 52.608 & 52.411 & 52.043 & 52.100 & 52.261 & 51.316 \\
\hline
FBC & 59.734 & 56.469 & 53.974 & 52.742 & 52.571 & 52.264 & 52.350 & NA     & 51.544  \\
\hline
\end{tabular}
\caption{The relaxed energy {per charge $E/Q$} of the $Q=\{1,2,\ldots,7\}$ Skyrmions on a $128\times 128\times 128$ lattice with $K_1=-1$, $K_2=0.248$ and $\kappa=0$. Note for $Q=6$: we {have found} a few extra stable local minima: the first two columns are results from merging, the 3rd column of PBC is a minimum relaxed from rational map, and the rational map with FBC is relaxed to $E/Q=52.264$.}
\label{table multi-charge}
\end{table}

The charge density isosurfaces displayed in Fig.~\ref{fig high charge} are found to be the same as in the Skyrme model for $Q=\{1,2,3\}$, and for $Q=4$ we start seeing small deviations~\cite{houghton1998rational}. For higher charge $Q=\{5,6\}$, the isosurfaces are found to have very different symmetries compared to the Skyrme model, but for $Q=7$ the isosurface is again only slightly distorted from the one in the Skyrme model. {One exceptional case} is a higher-energy $Q=6$ state we find with PBC, which {has the same charge density isosurface} as the lowest energy $Q=6$ solution in the Skyrme model. The reason for not seeing it with FBC is suspected to be that the energy barrier to other $Q=6$ states is very low, so it is easy to miss this solution {in the numerical relaxation}.

It is apparent from Fig.~\ref{fig high charge} that the lower charge solitons are qualitatively similar to the corresponding Skyrmions in the Skyrme model~\cite{BRAATEN1990147} as well BPS monopoles in $SU(2)$ Yang-Mills~\cite{hitchin1995symmetric}, both of which may be approximated by the rational map ansatz~\cite{houghton1998rational,battye1997symmetric,battye2002skyrmions},
\begin{align}
U(r,w) = \cos f(r)\,I+ i \sin f(r)n(w)\cdot\sigma.\label{def Rational map ansatz}
\end{align}
Here $n^i$ is a unit vector, which is a generalization of $n^i=x^i/r$ in the hedgehog ansatz \eqref{def Hedgehog}. The spatial coordinates $x^i=(x,y,z)$ are expressed in a spherical coordinate system $r,w$, where $w$ is a complex coordinate which is a function of the angles $\theta,\phi$,
\begin{align}
w\equiv\quad\tan \frac{\theta}{2}e^{i\phi}=\quad\frac{x+iy}{r+z}. \label{def W}
\end{align}
The dependence of $n$ on $w$ may be expressed in terms of an analytic function $R(w)$,
\begin{align}
n=\frac{1}{1+|R|^2}\left(2\,\text{Re} R, \,2\,\text{Im} R,\, 1-|R|^2\right).\label{eq n definition}
\end{align}
This function $R$ is the rational map from which the ansatz gets its name. It is a rational function $R=p/q$ where $p,q$ are polynomials with no common roots. The degree of $R$ is defined as the maximum degree of $p$ or $q$, and it turns out that the degree is simply equal to the baryon charge $Q$ of the ansatz $U$.

To determine how well the lower charge solitons found in the direct lattice simulation fit the rational map ansatz, the continuum description of the Hamiltonian \eqref{ham Continuum} was used to optimize the profile function $f$ given some rational map $R$. The quartic terms \eqref{lagr Quartic} in this case are a bit more complicated than the Skyrme model, where all integrals over angle either lead to an expression for the charge
\begin{align}
Q=\frac{r^2}{4\pi}\int d\Omega\,\frac{1}{2}\sum_\mu \left(\partial_\mu n\right)^2,\label{eq Q n integral}
\end{align}
or a single non-trivial integral $\mathcal{I}_0$,
\begin{align*}
\mathcal{I}_0\equiv \frac{r^4}{4\pi}\int  d\Omega\left[\frac{1}{2}\sum_\mu \left(\partial_\mu n\right)^2\right]^2.
\end{align*}
$\mathcal{I}_0$ may be easily expressed in terms of $R$ and minimized independently of the profile function~\cite{houghton1998rational}. On the other hand, the present model leads to four distinct angular integrals \eqref{def I} which are all coupled to the profile function and rather complicated if expressed in terms of $R$. In practice we simply took $R(w)$ to have the same discrete symmetry as it does in the Skyrme model, and for $Q\leq 4$ that completely fixes $R(w)$ so no minimization is necessary~\cite{houghton1998rational}. For $Q>4$, the parameters of the rational map were minimized directly in the lattice simulation as will be discussed below.

The structures of the Hamiltonian \eqref{ham Continuum} at $\kappa=0$ expressed in terms of the ansatz and averaged over solid angle are
\begin{gather}
\frac{1}{4\pi}\sum_\mu\int d\Omega 	\left(J^i_\mu\right)^2 =\left(f'\right)^2+\frac{2Q\sin^2 f}{r^2},\label{lagr Quad f}\\
	\frac{1}{4\pi}\sum_\mu\int d\Omega \left(	 \left(J^i_\mu\right)^2\right)^2=\frac{3}{5}\left(f^\prime\right)^4+\frac{2\sin^2 f}{r^2}\left(f^\prime\right)^2\mathcal{I}_1+\frac{\sin^4 f}{r^4}\mathcal{I}_2\label{lagr Quart J4 f}
\end{gather}
\begin{align}
	\frac{1}{4\pi}\sum_{\mu}\int d\Omega  \left(\partial_\mu J^i_\mu\right)^2&= \frac{3}{5}\left(f^{\prime\prime}\right)^2+\frac{4}{5r}f^\prime f^{\prime\prime}+\frac{8}{5r^2}\left(f^{\prime}\right)^2\non
	&\qquad +4\frac{\cos^2 f}{r^2}\left(f^{\prime}\right)^2\mathcal{I}_1+\frac{\sin^2 f}{r^4}\mathcal{I}_3-\frac{\sin^4 f}{r^4}\mathcal{I}_2\non
	&\qquad -2\frac{\cos f\, \sin f}{r^2}f^{\prime\prime}\mathcal{I}_1+2\frac{\cos f\, \sin f}{r^3}f^{\prime}\left(-2Q+\mathcal{I}_1+2\mathcal{I}_4\right),\label{lagr Quart dJdJ f}
\end{align}
with the integrals defined as,
\begin{align}
	\mathcal{I}_1&\equiv \frac{r^2}{4\pi}\sum_\mu\int d\Omega  \left(\left(x^\mu\right)^2 \left(\partial_\mu n^i\right)^2\right) \non%&\rightarrow \left(1-q\right)\non
	\mathcal{I}_2&\equiv \frac{r^4}{4\pi}\sum_\mu\int d\Omega\left(	 \left(\partial_\mu n^i\right)^2\right)^2 \non
		\mathcal{I}_3&\equiv \frac{r^4}{4\pi}\sum_\mu\int d\Omega \left(\partial^2_\mu n^i\right)^2\non
	\mathcal{I}_4&\equiv \frac{r^3}{4\pi}\sum_\mu\int d\Omega \left(x^\mu\,\partial_\mu n^i\partial^2_\mu n^i\right).\label{def I}
\end{align}
The values of the $\mathcal{I}$ integrals are given in Table \ref{table I integrals}, including the hedgehog special case $R(w)=w$, which was used in the previous section. It is seen that the energy of optimal rational map ansatz in the continuum comes fairly close to the energy of the more general low charge solitons in the lattice model.

\begin{table}
\begin{center}
\begin{tabular}{ c||c|c|c|c|c||c|c|c| }
		$Q$ & $R(w)$ & $\mathcal{I}_1$ & $\mathcal{I}_2$ & $\mathcal{I}_3$ & $\mathcal{I}_4$&  $E_{ansatz}/Q$ & $E/Q$ & $E/Q_{num}$ \\
		\hline
		$1$ & $w$ & $0.4$ & $1.6$ & $3.2$ & $-0.8$ & $59.81$ & $59.530$ & $61.308$\\
		$2$ & $w^2$ & $0.74926$ & $9.64307$ & $16.09$ & $-1.62537$& $59.92$ & $56.274$ & $57.687$	\\
		$3$ & $\frac{\pm i \sqrt{3} w^2-1}{w \left(w^2\mp i \sqrt{3}\right)}$ & $1.24335$ & $20.7566$ & $33.5133$ & $-2.37832$&  $57.92$ & $53.883$ & $55.189$	\\
		$4$ & $\frac{w^4+2 i \sqrt{3} w^2+1}{w^4-2 i \sqrt{3} w^2+1}$ & $1.97218$ & $30.2953$ & $48.4568$ & $-3.01391$&  $55.25$ & $52.608$ & $53.791$	\\
		%	$5$ & $\frac{w \left(w^4\pm 4 i w^2-3\right)}{-3 w^4\pm 4 i w^2+1}$ & $1.69048$ & $60.512$ & $94.9228$ & $-4.15476$	\\
		\hline
\end{tabular}
\caption{A comparison of the energy $E_{ansatz}$ of the rational map ansatz to the energy $E$ of the solitons found in the lattice simulation {with PBC}. All values are taken at $K_1=-1, K_2= 0.248$, {$\kappa=0$}, and the profile function was minimized in a finite volume of $r\leq 120$. $E$ is divided by both the exact charge $Q$ and a numerical charge $Q_{num}$ found from a finite difference approximation to \eqref{def Q Skyrme}.}\label{table I integrals}
\end{center}
\end{table}

%For Skyrmions with charge $Q>4$, while we may still use the previous method of ``merging'' to obtain them for the lattice model,
%the rational map ansatz provides \zw{an alternative way of producing initial spin configurations for a given charge $Q$. The energy of the optimal rational map is expected to be close to the fully relaxed ones.}
%According to Fig.~\ref{fig high charge}, the Skyrmions with large $Q$ may deviate from the spherically symmetric solution. In other words, the use of a profile function $f(r)$ with a single variable $r$ may not capture well the spatial anisotropies.
%Nevertheless, it is worth exploring if the rational map ansatz works well with the lattice model.

Unlike the situation
 in the Skyrme model, the profile function $f(r)$ and the rational function $R$ can not be minimized independently. {In this paper, we minimize them} simultaneously using a simulated annealing algorithm. Compared to local minimization, the simulated annealing method is advantageous in overcoming local minima with the help of thermal fluctuations. Typically we parametrize $f(r)$ by 20 to 30 discrete points and interpolate between them via Steffen's method which guarantees monotonicity. Our unit Monte Carlo (MC) step consists of updating $f(r)$ at each discrete point once, and updating each parameter in $R$ thirty times.
The initial temperature is $T_0=0.5 |K_1|$ and we bring it down to $T=0.001|K_1|$ in 3000 MC steps, then we use another 1000 MC steps for further equilibration at $T=0.001|K_1|$. We note that while the energy of the lattice model has to be evaluated at each MC update, there are only a few parameters to be minimized. This is in contrast to the full relaxation of the lattice model, where all $L^3$ quaternions $\bm{q}(x)$ are to be optimized.

{To illustrate these procedures, now} we consider the charge-5 Skyrmion of the lattice model. The rational map ansatz with $D_{2d}$ symmetry is~\cite{MantonBook}
\begin{equation}
R(w)= \frac{w (a+ibw^2+w^4)}{1+ibw^2+aw^4},\label{rational B5}
\end{equation}
where parameters $a$ and $b$ have to be optimized together with the profile function.

\begin{figure}[tbp]
\centering
\includegraphics[width=0.9\textwidth]{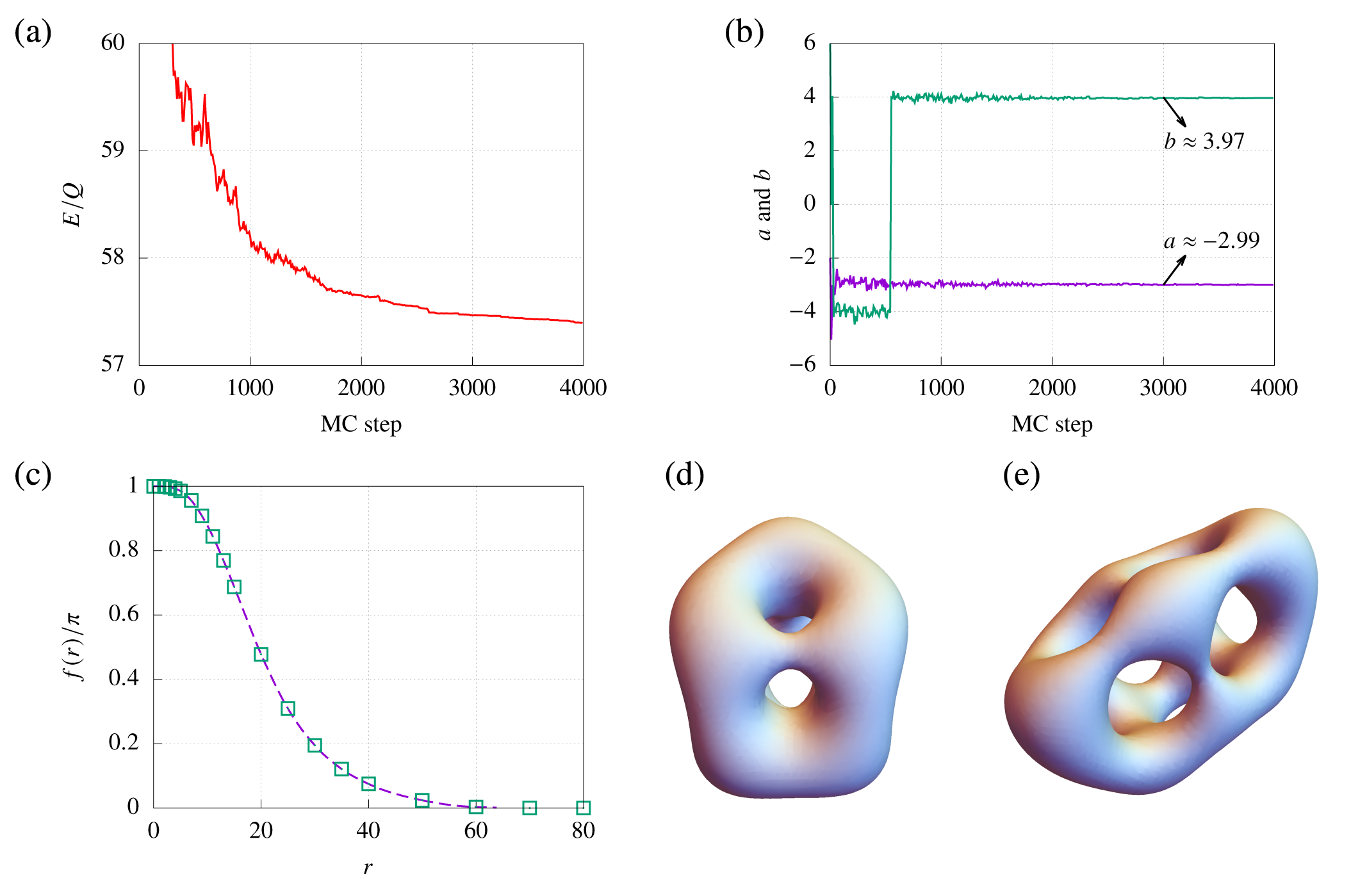}
\caption{Results of the charge-5 rational map ansatz~\eqref{rational B5} on a $L=128^3$ cubic lattice with $K_1=-1$, $K_2=0.248$, $\kappa=0$, and PBC. (a)(b) The evolution of energy and \{$a$, $b$\} during simulated annealing. (c)(d) The profile function and the charge density isosurface at the end of the simulated annealing (MC step$=4000$). The squares in (c) are the discrete points of $f(r)$ and the dashed line is the interpolation. (e) The charge density isosurface after fully relaxing the rational map ansatz.}
\label{fig rational B5}
\end{figure}

Figure~\ref{fig rational B5} shows the simulated annealing results of the charge-5 rational map. The optimized values $a\approx -2.99$ and $b \approx 3.97$ are quite close to the numbers of the Skyrme model ($a=-3.07$, $b=3.94$)~\cite{MantonBook}. Indeed, the charge density isosurface is the same as the one in the Skyrme model [Fig.~\ref{fig rational B5}(d)]. After further full relaxation, the energy and charge density isosurface converge to the results from merging [see Fig.~\ref{fig high charge}]. This result demonstrates both the usefulness and limitations of the rational map ansatz. On one hand, it allows us to construct Skyrmions with higher charge without going through the merging process. On the other hand, the rational map ansatz is sometimes incompatible with the anisotropies, and can become unstable towards the lower energy solution after full relaxation.

We have also performed simulated annealing on the $Q=6$ and $Q=7$ rational maps. For $Q=6$, we use the rational map with $D_{4d}$ symmetry:
\begin{equation}
	R(w) = \frac{w^4+ia}{w^2 (iaw^4+1)},
\end{equation}
where both the parameter $a$ and the profile function $f(r)$ are optimized. After further full relaxation, the result is found to be stable (the 3rd $Q=6$ plot in Fig.~\ref{fig high charge}) with PBC, while it tunnels to the lowest energy solution from merging (the 1st $Q=6$ plot in Fig.~\ref{fig high charge}) with FBC. The tunneling is suspected to be caused by the small energy barrier between different $Q=6$ states.

For $Q=7$, we use the rational map with $Y_h$ symmetry:
\begin{equation}
	R(w) = \frac{w^7-7w^5-7w^2-1}{w^7+7w^5-7w^2+1}.
\end{equation}
In this case, only the profile function $f(r)$ has to be optimized. The result of simulated annealing with full relaxation is found to also deviate slightly from the one in Skyrme model (Fig.~\ref{fig high charge}).

\begin{figure}[tbp]
\centering
\includegraphics[width=0.97\textwidth]{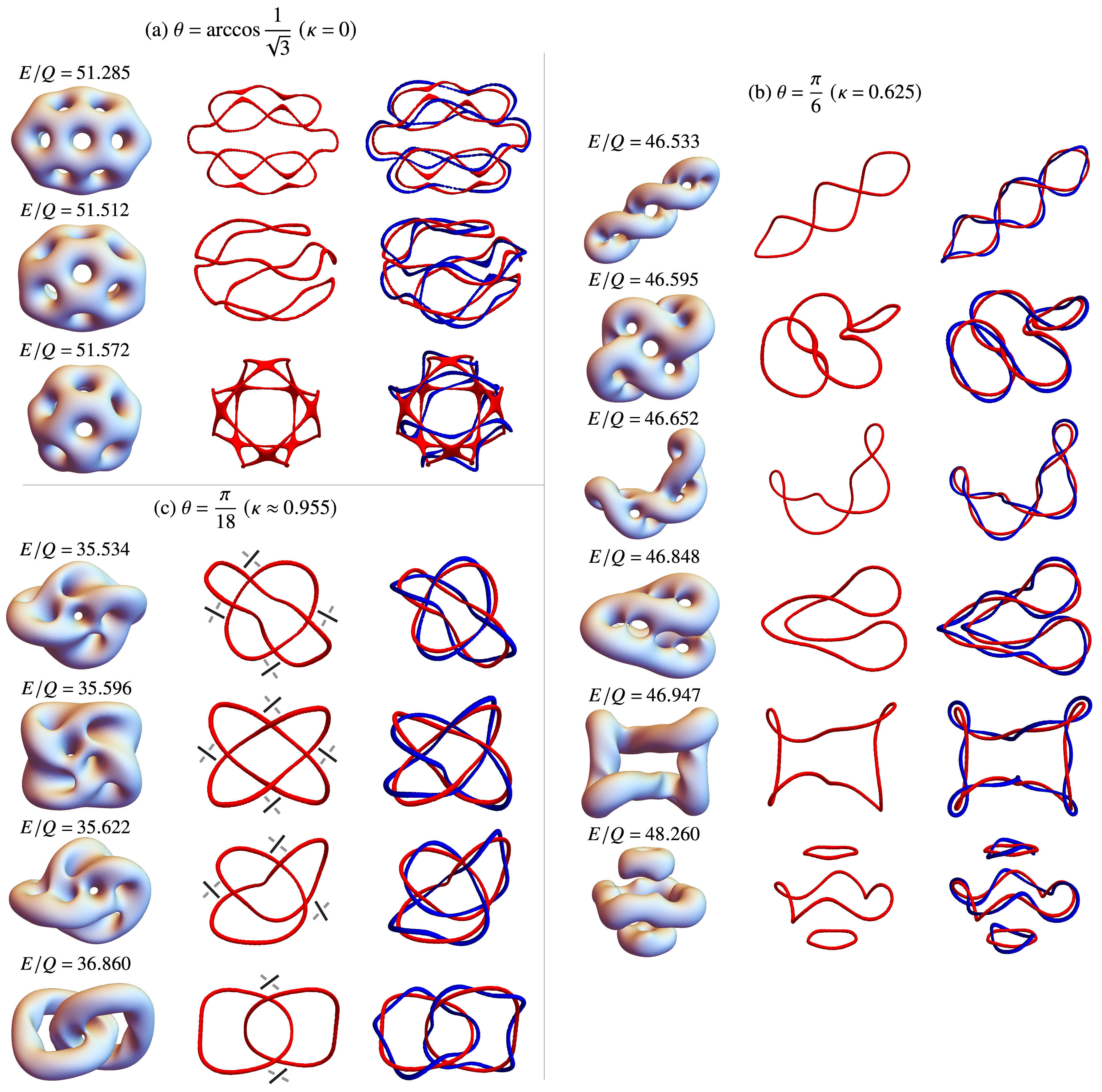}
\caption{The charge density isosurfaces and the position curves (red) and linking curves (blue), for fully relaxed charge $Q=10$ Skyrmion solutions on a $128\times 128\times 128$ cubic lattice with $K_1=-1$, $K_2=0.248$, $\kappa=\{ 0,\, 0.625,\,0.955 \}$, and FBC. The position curves in (a)(b) are all simple 1-rings except the bottom rows. The position curves in (c) include linked rings and knots, where the relative positions are illustrated by the crosses.}
\label{fig Q10}
\end{figure}

\subsection{Charge-10 Skyrmions and Position Curves} \label{sec charge-10}
{
In the $CP^1$ limit ($\beta = \kappa = 1$), the position curves were shown to have nontrivial structures including {linked rings} and knots~\cite{battye1998knots,sutcliffe2017skyrmion}. Here we show that such structures also appear naturally in our lattice model \eqref{ham S} when we interpolate to large $\kappa$.
}

%\zw{
%As we discussed, the position curves also carry the topological information of the underlying Skyrmion field. For instance, the linking structure of the two simple loops in Fig.~\ref{fig profile lattice charge1}(e) confirms that the charge is indeed $Q=1$ for the fully relaxed hedgehog configurations.
%}

{As we discussed in Sec.~\ref{sec Position curves and strings}, the position curves are not uniquely defined at the PCM limit ($\beta=\kappa=0$), due to the increased symmetry from $U(1)$ to $SU(2)$. In the quaternion representation, the energy \eqref{ham quaternion} is invariant under a global rotation \begin{equation}
\tilde{\bm{q}}(x)\equiv \bm{p} \bm{q}(x) \overline{\bm{p}}
\end{equation} for any given quaternion $\bm{p}$, {but this transformation does not leave the position curve invariant}. To get around this problem, in the following we always first find the optimal $\bm{p}$  that minimizes the energy for $\kappa \rightarrow 0+$ when plotting the position curves for $\kappa=0$.}

{Away from the $\kappa=0$ limit, both the energy and the third spin component $S^3(x)=q_0^2(x) + q_3^2(x)-q_1^2(x)-q_2^2(x)$ are invariant if $\bm{p}$ corresponds to rotation around the $\hat{z}$ axis. In other words, the position curve defined by ${S}=(0,0,-1)$ is unique under such global transformations, but position curves along other spin directions are not generally invariant. For this reason, we mainly consider the ${S}=(0,0,-1)$ position curves in this section; while the linking curves at ${S}=(0,-1/\sqrt{2},-1/\sqrt{2})$ are not unique, we also plot them together with the position curves to show their relative linking structures.}

{Figure~\ref{fig Q10} shows the fully relaxed $Q=10$ solutions of the lattice model for three different choices of $\kappa$. For each $\kappa$, since we used multiple ways of ``merging'' and rational maps, multiple local minima are obtained. The lowest energy $\kappa =0$ solutions [Fig.~\ref{fig Q10}(a), first two rows] shows clear deviation from the relaxed rational map with $D_{4d}$ symmetry [Fig.~\ref{fig Q10}(a), bottom row]. The position curves are all simple 1-rings except the relaxed rational map in which the position curve forms a net. We note that such net-like structure was also observed for low charge Skyrmions at $\beta=0$ in the Skyrme model~\cite{ward2004skyrmions}. }

{Both the topological charge isosurfaces and the position curves change dramatically as $\kappa$ is tuned away from zero. For $\kappa=0.625$, the position curves are found to be either a simple 1-ring, or several 1-rings which are disjointed [see Fig.~\ref{fig Q10}(b)]. For a large value $\kappa\approx 0.955$, the position curves start developing nontrivial topological features, including both linked rings and knots [see Fig.~\ref{fig Q10}(c)].
This is indeed as expected since such topological structures were known to exist in the $\kappa=1$ limit~\cite{sutcliffe2017skyrmion}.}

\section{Conclusion and discussion} \label{sec Conclusion}

{3D Skyrmions proper in the sense of the Skyrme model are shown to be stabilized in a frustrated spin model on the cubic lattice.
By tuning a parameter that describes the ``collinearity'' of the magnetic ground state, {the model interpolates between two limits with $S^3$ and $S^2$ target spaces}. In the $S^3$ limit, the Skyrmion solutions are found to be qualitatively the same as in the Skyrme model for small $Q$, and they start to deviate for $Q\gsim 4$. Near the $S^2$ limit, the position curves of the Skyrmions are found to develop nontrivial topological structures including {linked rings} and knots.
}

{Since the lattice model considered in this paper can be regarded as the low-energy effective model for a broad class of 3D non-collinear magnets where $SO(3)$ rotation is the only low-energy mode, it is expected that 3D Skyrmions should also appear in realistic models that have {SO(3) Goldstone modes, whether squashed or not.} %a $\pi_3(S^3)$ (squashed or not) low-energy description.
	To this end, it is worth emphasizing}
%To date, 3D Skyrmions (Skyrmion defects or Skyrmion crystals) have not been discovered in condensed matter experiments. Here we discuss
a few necessary ingredients in the search of 3D Skyrmion excitations (defects) in magnetic systems~\cite{BatistaEtAl2018}. First, a non-collinear ground state is required to ensure a target space homeomorphic to $S^3$, which can be commonly realized in frustrated spin systems. Second, the Skyrmion size has to be much larger than the lattice spacing for the excitation to be topologically protected (energy barrier {large enough} between different topological sectors). For the toy model considered in this paper, the Skyrmion size becomes large when $K_2/|K_1|\lesssim 1/4$. More generally, this condition is satisfied for systems near a Lifshitz point (a commensurate to incommensurate transition).

{Similar to the 2D Skyrmion crystals that are commonly studied in condensed matter systems,}
3D Skyrmion crystals are expected to be realized {as the ground state (vacuum of the theory)} on the other side of the Lifshitz transtion.
{We note that the precise definition of ``Skyrmion crystal'' is slightly different in the condensed matter and high-energy literatures: the condensed matter community often refers ``Skyrmion crystal'' as the $T=0$ ground state or the finite-$T$ equilibrium state (Skyrmion crystal becomes the new vacuum), while in high-energy ``Skyrmion crystal'' is often referred as an excited state in the original vacuum.
In both cases, the Skyrmion crystal can be described as ``multi-$\bm{Q}$'' states (linear combination of multiple incommensurate spirals), whose energy can be quite close to other single- or multi-$\bm{Q}$ states.}
Such degeneracy {is often} lifted by spin anisotropy, magnetic field, thermal and {quantum} fluctuations.
{Besides magnetic systems, we note that 3D Skyrmion crystals were also predicted to be realized in cold atom systems described by multicomponent imbalanced superfluids~\cite{Samoilenka2020}.}

{Lorentz transmission electron microscopy is often used for direct visualization of 2D magnetic Skyrmions.}
More recently, magnetic
X-ray tomography was successfully applied to 3D systems for the visualization of Skyrmion strings and Hopfions~\cite{seki2021direct}. In principle, 3D Skyrmions could also be detected by the same X-ray tomography methods. As we noted in this paper, the close connection between 3D Skyrmions and Hopfions implies that visualization of
%Hopfions
{the underlying position curves}
{can be}
{strong evidence of 3D Skyrmion formation, if the ground state is known to be non-collinear.}
%first step towards realizing the underlying 3D Skyrmion structure.
%On the other side of the Lifshitz point,
{Small angle neutron scattering is also a useful tool to see the multi-$\bm{Q}$ structure of the underlying spin arrangements, which serves as indirect evidence of Skyrmion crystal formation.}
%Skyrmion crystals (2D or 3D) are multi-$Q$ incommensurate structures. When direct visualization techniques are not available, . Generally, multi-$Q$ peaks (from a single domain) in the static spin structure factor are expected to be seen if Skyrmion crystals are realized as the ground state.

{Finally, let us note that the picture we have presented of the continuity between Skyrmions and string-like Hopfions may have some relevance in high-energy physics to the study of the Skyrme model and its various extensions and modifications. As we have discussed in Sec. \ref{sec Position curves and strings} a solution with non-zero baryon charge in the Skyrme model may equivalently be considered as a knotted or twisted loop of string with long range interactions associated with the $J^3$ field that winds around the string core. It may be difficult to make use of this picture in a concrete way since for minimum energy configurations the radius of curvature of the position curve is on the same order of magnitude as the string thickness. But at the very least in the limit of the Faddeev-Niemi model there are unstable configurations which are well described by a thin string ansatz along these lines \cite{battye1999solitons}.
	
	The main qualitative difference between the Faddeev-Niemi model at $\beta=1$ and the squashed Skyrme model for $\beta<1$ in this point of view is that former involves local strings whereas the latter involves global strings with long range interactions. It may be interesting to explore whether this is connected to the difference which must be present in large $Q$ solutions given the linear energy bound \eqref{eq BPS bound} we have found here. While these considerations are certainly more speculative than the possibility of direct detection of 3D Skyrmions in condensed matter systems discussed above, it may also be fruitful to investigate these analogies between the Skyrme model and systems of stringy topological defects in further detail.} 

%\zw{[xx I think the following paragraph can be removed. See discussion in the 2nd paragraph in this section. xx]}
%While the lattice model considered in this paper is a bit artificial, it bears the minimal ingredients and demonstrates the possibility of realizing the 3D Skyrmion in magnetic systems. For more realistic (complex) models, e.g. the pyrochlore lattice, it is expected that 3D Skyrmions should also be realized near the Lifshitz point.

\begin{acknowledgments}
	We would like to acknowledge Cristian Batista for helpful discussions. D.S. was supported in part by the U of MN Doctoral Dissertation Fellowship. During the writing of this paper, Z.W. was supported by the U.S. Department of Energy through the University of Minnesota Center for Quantum Materials, under Award No. DE-SC-0016371. C.N. was supported by the Olle Engkvist foundation, Grant No 204-0185. M.S. is supported in part by DOE Grant No. DE- SC0011842.
\end{acknowledgments}

\end{document}